%% file: oblique.tex
\PassOptionsToPackage{unicode}{hyperref}
\PassOptionsToPackage{hyphens}{url}
\documentclass[
  astrosymb, twocolumn, twocolappendix, tighten]{aastex631}
\usepackage{xcolor}
\usepackage{amsmath,amssymb}

\usepackage{iftex}
\ifPDFTeX
  \usepackage[T1]{fontenc}
  \usepackage[utf8]{inputenc}
  \usepackage{textcomp} 
\else 
  \usepackage{unicode-math} 
  \defaultfontfeatures{Scale=MatchLowercase}
  \defaultfontfeatures[\rmfamily]{Ligatures=TeX,Scale=1}
\fi
\usepackage{lmodern}
\ifPDFTeX\else
\fi
\IfFileExists{upquote.sty}{\usepackage{upquote}}{}
\IfFileExists{microtype.sty}{
  \usepackage[]{microtype}
  \UseMicrotypeSet[protrusion]{basicmath} 
}{}
\makeatletter
\@ifundefined{KOMAClassName}{
  \IfFileExists{parskip.sty}{%
    \usepackage{parskip}
  }{
    \setlength{\parindent}{0pt}
    \setlength{\parskip}{6pt plus 2pt minus 1pt}}
}{
  \KOMAoptions{parskip=half}}
\makeatother
\usepackage{longtable,booktabs,array}
\usepackage{calc} 
\usepackage{etoolbox}
\makeatletter
\patchcmd\longtable{\par}{\if@noskipsec\mbox{}\fi\par}{}{}
\makeatother
\IfFileExists{footnotehyper.sty}{\usepackage{footnotehyper}}{\usepackage{footnote}}
\makesavenoteenv{longtable}
\usepackage{graphicx}
\makeatletter
\newsavebox\pandoc@box
\newcommand*\pandocbounded[1]{
  \sbox\pandoc@box{#1}%
  \Gscale@div\@tempa{\textheight}{\dimexpr\ht\pandoc@box+\dp\pandoc@box\relax}%
  \Gscale@div\@tempb{\linewidth}{\wd\pandoc@box}%
  \ifdim\@tempb\p@<\@tempa\p@\let\@tempa\@tempb\fi
  \ifdim\@tempa\p@<\p@\scalebox{\@tempa}{\usebox\pandoc@box}%
  \else\usebox{\pandoc@box}%
  \fi%
}
\def\fps@figure{htbp}
\makeatother
\providecommand{\tightlist}{%
  \setlength{\itemsep}{0pt}\setlength{\parskip}{0pt}}
\usepackage[]{natbib}
\usepackage{mathptmx,txfonts,tikz,bm}
\usepackage[capitalize]{cleveref}

\newcommand{\annotate}[2]{\begin{tikzpicture}
    \node[anchor=south west,inner sep=0,align=center] (image) at (0,0) {
    #1
    };
    \begin{scope}[x={(image.south east)},y={(image.north west)}]
    #2
    \end{scope}
\end{tikzpicture}}

\renewcommand{\edit}[2]{{\ifnum#1 <  1 %
#2%
\else%
\textbf{#2}%
\fi}}

\input{macros.tex}

\usepackage[]{natbib}
\usepackage{bookmark}
\IfFileExists{xurl.sty}{\usepackage{xurl}}{} 
\received{October 1, 2024}
\revised{December 26, 2024}
\accepted{December 26, 2024}
\submitjournal{The Astrophysical Journal}

\let\[\relax \let\]\relax 
\DeclareRobustCommand{\[}{\begin{equation}}
\DeclareRobustCommand{\]}{\end{equation}}

\begin{document}

\title{Signatures of Core-Envelope Rotational Misalignment in the Mixed-Mode Asteroseismology of Kepler-56}

\correspondingauthor{Joel Ong}
\email{joelong@hawaii.edu}

\author[0000-0001-7664-648X]{J. M. Joel Ong \chinesename}
\altaffiliation{NASA Hubble Fellow}
\affiliation{Institute for Astronomy, University of Hawaiʻi, 2680 Woodlawn Drive, Honolulu, HI 96822, USA}

\shortauthors{Ong}
\shorttitle{Misaligned Mixed Modes and Kepler-56}
\begin{abstract}
Existing asteroseismic rotational measurements assume that stars rotate around a single axis. However, tidal torques from misaligned companions, or their possible engulfment, may bring the rotational axis of a star's envelope out of alignment with its core, breaking azimuthal symmetry. I derive perturbative expressions for asteroseismic signatures of such hitherto unexamined rotational configurations, under the ``shellular approximation'' of constant rotation rates on radially stratified mass shells. In the aligned case, the distribution of power between multiplet components is determined by the inclination of the rotational axis; radial differential misalignment causes this to vary from multiplet to multiplet. I examine in particular detail the phenomenology of gravitoacoustic mixed modes as seen in evolved sub- and red giants, where near-resonance avoided crossings may break geometrical degeneracies. Upon applying the revised asteroseismic observational methodology that results from this theoretical discussion to revisit Kepler-56 --- a red giant with a misaligned planetary system --- I find that its core and envelope rotate around different rotational axes. While the rotational axis of its core is indeed misaligned from the orbit normal of its transiting planets (consistently with earlier studies), its envelope's rotational axis is close to lying in the sky plane, and may well be aligned with them. More detailed asteroseismic modelling, and spectroscopic follow-up, will be required to fully elucidate the full spin-orbit geometry of the Kepler-56 system, and potentially discriminate between hypotheses for how it formed.
\end{abstract}
\keywords{Asteroseismology (73), Exoplanet formation (492), Exoplanets (498), Hot Jupiters (753), Planetary alignment (1243), Red giant stars (1372), Stellar oscillations (1617), Stellar rotation (1629), Star-planet interactions (2177), Theoretical techniques (2093)}

\def\sectionautorefname{Section}
\def\subsectionautorefname{Section}
\def\subsubsectionautorefname{Section}

\section{Introduction}\label{introduction}

As stars ascend the red giant branch after main-sequence evolution, their radiative cores contract, and convective envelopes expand. Standard assumptions about angular momentum conservation and transport in stellar interiors suggest that this causes their surface rotation rates to become almost observationally negligible --- an assumption that is empirically largely bourne out \citep[e.g.][]{li_asteroseismic_2024}. However, a nontrivial fraction of the \Kepler~sample of pulsating red giants does exhibit rapid surface rotation \citep[e.g.][]{ceillier_surface_2017}, and of these, only 15\% are known to possess stellar-mass orbital companions that might have spun them up tidally \citep{gaulme_active_2020}.

One proposed explanation for the existence of the rest is that planetary companions could have spun up their envelopes through the action of tidal torques. Both single-target \citep[e.g.][]{yee_orbit_2020, vissapragada_tidal_2022} and population \citep{saunders_obliquity_2024} studies suggest the action of such tidal torques when detectable planetary companions are present. Alternatively, former companions may have deposited their orbital angular momentum into the stellar envelope directly during engulfment by the star. Indeed, recent observations may have captured stars during \citep[e.g.][]{de_infrared_2023} and shortly after \citep[e.g.][]{ong_zvrk_2024} such engulfment events, leaving behind rapid rotation as an engulfment signature.

Population studies of exoplanets around main-sequence stars \citep[e.g.][]{winn_hot_2010, winn_occurence_2015, albrecht_preponderance_2021} suggest a nontrivial population of highly-inclined companion orbits --- i.e.~with large spin-orbit misalignments --- particularly around hot stars. However, these highly-inclined orbits appear to be less common around evolved stars, suggesting realignment occurring after the main sequence \citep{saunders_obliquity_2024}.

As with tidal spin-up, this post-main-sequence realignment may operate by tidal torques acting on tidal bulges raised on convective envelopes \citep[as suggested in e.g.][]{lai_tidal_2012, rogers_lin_2013}, similarly to the mechanics of tidal spin-up \citep[e.g.][]{brown_discrepancies_2014, maxted_comparison_2015, arevalo_further_2021}. Should this realignment of the envelope happen faster than the envelope can torque the core, then the two may be brought out of alignment, at least temporarily. Conversely, stellar spin-up through engulfment may occur in initially misaligned systems as well. Planetary engulfments are more probable with highly eccentric companions than those in circular orbits \citep[e.g.][]{stephan_destroyers_2018, stephan_giant_2021}; such high-eccentricity configurations, being originally brought out of circularisation by multibody interactions, are potentially also highly inclined. Again, the deposition of misaligned orbital angular momentum into the stellar envelope would at least momentarily bring the envelope's rotational axis out of alignment with that of the core.

Since single-star evolution is not thought to naturally produce such rotational misalignment, the ability to diagnose and characterise it observationally would permit such historical or ongoing companion interactions to be quickly identified. Asteroseismology may provide this capability, as it is our only direct observational probe of these internal, rather than surface, rotational properties of stars. However, existing observational prescriptions for seismic rotational measurement assume a priori that the star has only one rotation axis everywhere. Generalising this, we present a misaligned-pulsator model including stratification of the direction of the rotational axis (\autoref{sec:construction}). As a result of this development, we propose the use of varying amplitude ratios between different rotationally-split multiplets as an observational diagnostic signature for any such potential internal misalignment.

We will then apply this construction to a simple two-zone model of radial differential rotation, of the kind ordinarily studied using gravitoacoustic mixed modes in evolved solar-like oscillators (\autoref{sec:twozone}). The widely-used Jeffreys-Wentzel-Kramers-Brillouin \citep[JWKB, e.g.][]{gough_jwkb_2007} asymptotic description of these mixed modes assumes from the outset that the preferred pulsational axes of normal modes in the core and envelope are aligned --- by performing separation of variables before any inspection of the radial problem --- even in the presence of multiple mode cavities \citep{shibahashi_modal_1979, unno_nonradial_1989, gough_rotation_1990, gough_linear_1993}. By contrast, our analytic description of the two-zone model in such misaligned pulsators permits the two families of p- and g-modes to separately pulsate along different preferred axes --- which need not necessarily be aligned --- while still coupling to each other. We explore the phenomenology of this construction, and provide quantitative prescriptions for the diagnosis of internal misalignment from mixed-mode frequencies and amplitudes.

Our results suggest that such misalignment may already have been unknowingly observed in the field. In \autoref{sec:k56}, we examine a case study of this --- the multi-planet host star, Kepler-56. Our findings suggest that the rotational axis of its envelope points in a different direction --- potentially closer to alignment with the orbits of its transiting planets --- than the rotational axis of its core. Follow-up measurements of the Rossiter-McLaughlin effect \citep{rossiter_detection_1924, mclaughlin_results_1924}, to constrain the obliquity of its transiting planets relative to its \emph{surface} rotation, may yield more insight into the nature of this system, and the physical processes associated with potential core-envelope or spin-orbit realignment. We conclude (in \autoref{sec:discussion}) with some discussion about the broader implications of both our findings regarding Kepler-56 in particular, and of our newfound ability to inspect internal rotational misalignment in general.

\section{Analytic Construction}\label{sec:construction}

Normal modes of oscillation in slowly rotating stars are indexed by three integers \(n, \ell, m\), where modes indexed by the same radial order \(n\) and latitudinal degree \(\ell\), but different azimuthal order \(m\), have (to leading order) identical radial dependences, and horizontal dependences specified by the spherical harmonic functions \(Y^\ell_m(\theta, \varphi)\), as constructed with respect to some notional coordinate system. In the absence of rotation, these modes of the same \(\ell\) and \(n\) but different \(m\) pulsate at identical mode frequencies, with \(2\ell + 1\) such degenerate modes for each combination of \(\ell\) and \(n\). Slow rotation around the \(z\)-axis induces a frequency perturbation into each mode proportional to \(m\), breaking this degeneracy.

Existing treatments of oblique pulsations, e.g.~as developed in \citet{shibahashi_rotational_1985}, consider the case of a single pulsation axis misaligned from a single rotation axis, where the pulsation axis is itself assumed to rotate around this rotational axis in the observer's stationary reference frame. As a result of this parametric time dependence of the pulsation axis, rotational multiplets in such a configuration may exhibit further hyperfine splitting, producing up to \((2\ell+1)^2\) multiplet components \citep[e.g.][]{gough_rotation_1990}. By contrast, if the core and envelope of a star should be set rotating along differently aligned axes, the core's rotational axis does not rotate around the envelope's, nor vice versa. While the two will exert torques on each other through various physical mechanisms \citep{aerts_angular_2019}, pulsations occur on much shorter timescales than those for angular momentum transport. Thus, at least phenomenologically, each rotational axis can be assumed to be fixed in the observer's stationary frame, yielding only \(2\ell+1\) multiplet components (neglecting the effects of magnetism).

Our treatment of oblique pulsations to describe such misaligned configurations therefore relies on a different mathematical construction, which in turn produces different phenomenology, from this existing oblique-pulsator model. Nonetheless, its fundamental building blocks are identical. In this section, we first remind the reader of some well-known properties of the spherical harmonics and associated spin matrices (\autoref{sec:matrices}), and then of existing applications of this analytic formalism to the asteroseismology of slow rotators (\autoref{sec:rotation}). We shall choose notation in this exposition so that, when we turn our attention to potential misalignment at the end of this section (\autoref{sec:misalignment}), our generalised expressions remain compact, with minimal differences from the aligned case.

\subsection{Notation: Rotation and Angular Momentum Matrices}\label{sec:matrices}

Our development here draws heavily upon the existing literature both pertaining to the phenomenology of slow rotation in asteroseismology \citep[e.g.][]{lyndenbell_stability_1967, gough_rotation_1990, aertsbook}, as well as more generally regarding the appearance of similar matrices in the quantum mechanics of angular momentum \citep[e.g.][]{landau_quantum_1965}. As a matter of terminology (following standard nomenclature in physics), we will refer to matrices that transform spherical harmonics in one coordinate system to linear combinations of those in a rotated coordinate system as ``rotation matrices''. Conversely, we will refer to matrix representations (expressed in the basis of spherical harmonics) of linear perturbations to the wave operator arising from stellar rotation as ``angular momentum matrices''. The two are interrelated: the transformation of the angular momentum matrices between rotated coordinate systems is specified by rotation matrices, while the rotation matrices are generated by exponentiating angular momentum matrices.

Under a coordinate rotation --- e.g.~such that a unit vector pointing along the \(z\)-axis is rotated to point in the direction \(\hat{\mathbf{n}}\) --- each spherical harmonic in the rotated coordinate system may be expressed as linear combinations of only spherical harmonics in the original coordinate system of the same \(\ell\):
\[
\begin{aligned}
    Y^\ell_m(\theta_{\hat{\mathbf{n}}}, \varphi_{\hat{\mathbf{n}}}) &= \sum_{\ell', m'}\left<Y^\ell_m(\theta_{\hat{\mathbf{n}}}, \varphi_{\hat{\mathbf{n}}}), Y^{\ell'}_{m'}(\theta, \varphi)\right> Y^{\ell'}_{m'}(\theta, \varphi) \\
    &\equiv \sum_{m'} D^\ell_{m', m} Y^\ell_{m'}(\theta, \varphi). \label{eq:dmatrix}
\end{aligned}
\]
This being the case, spherical harmonics of degree \(\ell\) remain orthogonal to those of degree \(\ell' \ne \ell\) under coordinate rotations, so for the remainder of this paper, we will accordingly neglect coupling between different \(\ell\).

These rotation matrices \(\mathbf{D}^\ell\), of rank \(2\ell+1\), are known as Wigner's D-matrices. They are unitary, as the inner product is preserved under rotations, and therefore must be expressible in the generic form \(\mathbf{D}^\ell = \exp \left[\mathrm{i} \phi \mathbf{J}\right]\), relating them to rotation angles \(\phi\), and traceless Hermitian matrices \(\mathbf{J}\), under the exponential map. These \(\mathbf{J}\) represent the angular momentum operator along the axis of rotation.

We now recount some well-known properties of Wigner's D-matrices, which are elements of the irreducible representation of the 3D rotation group SO(3) of rank \(2\ell + 1\). When the pulsation axis (i.e.~the \(z\)-axis of the coordinate system in which one constructs spherical harmonics) is chosen to coincide with the rotation axis of a star, the rotating normal modes are also specified by the basis spherical harmonics in the perturbative limit of slow rotation, and so the matrix elements \(\left<Y^\ell_m, \hat{\mathcal{R}}Y^\ell_{m'}\right>\) of the perturbation to mode frequencies also yield a diagonal matrix, with elements proportional to \(m\). These matrix elements are proportional to the matrix representation of the angular momentum operator along the \(z\)-axis, \(\mathbf{J}_z\), which takes the form
\[
    \mathbf{J}_z = \mathrm{diag}(-\ell, -\ell + 1, \ldots, \ell - 1, \ell), \label{eq:jz}
\]
so that \(\mathbf{D} = \exp[\ii \mathbf{J}_z \phi]\) has matrix elements \(D_{m,m'} = \delta_{m,m'}e^{\ii m \phi}\).

Because the D-matrices \(\mathbf{D}^\ell\) are representations of SO(3) specifically, there must be, addition to \(\mathbf{J}_z\), two additional special Hermitian matrices \(\mathbf{J}_x\) and \(\mathbf{J}_y\), of the same rank as \(\mathbf{J}_z\), that generate each representation of SO(3). These each have the same eigenvalues, and satisfy the commutation relations
\[
    [\mathbf{J}_i, \mathbf{J}_j] = \ii \epsilon_{ijk} \mathbf{J}_k,
\]
where \(\epsilon_{ijk}\) is Levi-Civita's totally antisymmetric symbol. Moreover,
\[
    \mathbf{J}_x^2 + \mathbf{J}_y^2 + \mathbf{J}_z^2 = \ell(\ell + 1) \mathbb{I}_{2\ell + 1}.
\]
For \(\ell = 1\), which features heavily in this work, we have in particular that
\[
\mathbf{J}_x = {1\over\sqrt2}\begin{bmatrix}0&1&0\\1&0&1\\0&1&0\end{bmatrix},\ 
\mathbf{J}_y = {1\over\sqrt2}\begin{bmatrix}0&\ii&0\\-\ii&0&\ii\\0&-\ii&0\end{bmatrix},\ 
\mathbf{J}_z = \begin{bmatrix}-1&0&0\\0&0&0\\0&0&1\end{bmatrix}.
\]

Each \(\mathbf{D}^\ell = \exp \left[\mathrm{i} \phi \mathbf{J}\right]\) corresponds to rotation by an angle \(\phi\), around an axis specified by \(\mathbf{J}\). In particular, rotations around the \(y\)-axis of some notional reference coordinate system by an angle \(\beta\), which map \(\mathbf{e}_z\) to the unit vector
\[
    \hat{\mathbf{u}}(\beta) = \cos \beta\ \mathbf{e}_z + \sin \beta\ \mathbf{e}_x,\label{eq:rotate}
\]
also generate D-matrices of the special form
\[
    \mathbf{d}^\ell = \exp\left[\ii \beta \mathbf{J}_y\right],
\]
often referred to as Wigner's (small) d-matrix. For \(\ell = 1\),
\[
\mathbf{d}^{\ell = 1}\left(\beta\right) = \begin{bmatrix}
 \cos^2 {\beta\over2} & \sqrt{2}\cos{\beta\over2}\sin{\beta\over2} & \sin^2 {\beta\over2} \\
 - \sqrt{2}\cos{\beta\over2}\sin{\beta\over2} & \cos\beta &  \sqrt{2}\cos{\beta\over2}\sin{\beta\over2} \\
 \sin^2 {\beta\over2} & -\sqrt{2}\cos{\beta\over2}\sin{\beta\over2} & \cos^2 {\beta\over2} \end{bmatrix}.\label{eq:dipoledmatrix}
\]
In general, any D-matrix may be factorised into a product of rotations by three Euler angles. We adopt the ``\(z\)-\(y\)-\(z\)'' convention for this factorisation as
\[
\mathbf{D}^\ell(\alpha, \beta, \gamma) = \exp\left[\ii \alpha \mathbf{J}_z\right] \mathbf{d}^{\ell}(\beta) \exp\left[\ii \gamma \mathbf{J}_z\right], \label{eq:euler}
\]
We may equivalently use these expressions, with rotations of opposite sign composed in opposite order, to describe rotations of the angular momentum vector (and thus operator) itself, keeping the coordinate system fixed.

Angular momentum operators transform under rotations between coordinate systems by the map \(\mathbf{J} \mapsto {\mathbf{D}^\ell}^\dagger \mathbf{J} \mathbf{D}^\ell\). However, since \(\mathbf{D}^\ell\) is unitary, the eigenvalues of \(\mathbf{J}\) remain invariant. In particular, this transformation maps each basis angular-momentum matrix to a linear combination of them, \(\mathbf{J}_i \mapsto \sum_{j} A_{ji} \mathbf{J}_j\), as specified by the adjoint representation \(\mathbf{A}\) --- the same one applied to the basis vectors themselves by the coordinate rotation as \(\mathbf{e}_i \mapsto \sum_{j}\mathbf{A}_{ji}\mathbf{e}_j\). Therefore, it maps each linear combination of these basis matrices to a different linear combination of the same basis matrices. To more conveniently discuss this, we shall use the shorthand notation
\[
v_x \mathbf{J}_x + v_y \mathbf{J}_y + v_z \mathbf{J}_z \equiv \hat{\mathbf{v}} \cdot \vec{\mathbf{J}},\label{eq:ndotJ}
\]
to denote a map from vectors \(\mathbf{v} = \sum_i v_i \mathbf{e}_i\) to the associated linear combinations of basis spin matrices. In this notation we may state a well-known identity of how the angular momentum vectors are transformed by such rotation:
\[\mathbf{v} \mapsto \mathbf{Av} \iff \left(\mathbf{v}\cdot\vec{\mathbf{J}}\right) \mapsto {\mathbf{D}^\ell}^\dagger\left(\mathbf{v}\cdot\vec{\mathbf{J}}\right) {\mathbf{D}^\ell} = \left((\mathbf{Av})\cdot\vec{\mathbf{J}}\right).\]
For example, for rotations around the y-axis as in \cref{eq:rotate},
\[
    {\mathbf{d}^\ell}^\dagger \mathbf{J}_z \mathbf{d}^\ell = \cos \beta\ \mathbf{J}_z - \sin \beta\ \mathbf{J}_x \equiv \hat{\mathbf{u}}(-\beta) \cdot \vec{\mathbf{J}}.
\]
More generally, every \(\mathbf{D}^\ell\) is associated with a rotation that maps the basis vector \(\mathbf{e}_z\) to some unit vector \(\hat{\mathbf{n}}\), whereby
\[
    \mathbf{e}_z \mapsto \hat{\mathbf{n}} \implies \mathbf{J}_z \mapsto {\mathbf{D}^\ell}^\dagger \mathbf{J}_z \mathbf{D}^\ell = \hat{\mathbf{n}}\cdot\vec{\mathbf{J}}.
\]

\subsection{Notation: Combining Rotation and Pulsations}\label{sec:rotation}

We now recount some standard results in the ``shellular rotation'' approximation of radial differential rotation, where the star is decomposed into concentric spherical mass shells indexed by the physical radial coordinate \(r\), associated with each of which is a rotational angular frequency \(\Omega(r)\). We further restrict the analysis in this work to ignore latitudinal differential rotation --- i.e.~we model the rotational motion of each concentric mass shell as being constant. At this level of approximation, for a multiplet of modes with nonrotating frequency \(\omega_{n\ell}\), when all mass shells are set rotating around a single rotational axis, the resulting rotating mode frequencies are known to be given at leading order in \(\Omega\) by
\[
    \omega_{n\ell m} \sim \omega_{n\ell} + m R_{n\ell, n\ell} + \mathcal{O}(\Omega^2)\equiv \omega_{n\ell} + m \delta\omega_{n\ell}.\label{eq:sym}
\]
Here \(R_{n\ell, n\ell}\) is the diagonal element of a matrix specified by the mode eigenfunctions and rotational profile \(\Omega(r)\). Restricting our attention to a single \(\ell\) henceforth, we recall that
\[
\begin{aligned}
    R_{n,n'} &=  \int \left\{\Omega(r)\  r^2 \rho_0 \left(\xi_{r, n} \xi_{r, n'}+ [\ell(\ell+1) - 1] \xi_{t, n} \xi_{t, n'}\right.\right. \\ &\left.\left.\hspace{4em} - \xi_{r, n} \xi_{t, n'} - \xi_{t, n} \xi_{r, n'}\right)\right\} \mathrm d r\\
    &\equiv B_{n,n'} \int \mathrm d r\ \Omega(r) K_{n,n'}(r),\label{eq:rotkernel}
\end{aligned}
\]
where we have defined overall sensitivity constants \(B_{n,n'}\)\footnote{these are sometimes also denoted \(\beta_{ij}\) in the seismic literature, but to avoid ambiguity we will use \(\beta\) in this work only for Euler angles describing misalignment between coordinate systems or angular momentum vectors.} so that the kernel functions on the diagonal \(K_{nn}\) are each of unit integral, and off the diagonal \(B_{n,n'} \equiv \sqrt{B_{nn} B_{n',n'}}\). More generally, the rotating mode frequencies over the basis set of different radial eigenfunctions, indexed by \(n\) and \(\ell\), emerge in this aligned configuration as solutions to the quadratic Hermitian eigenvalue problem \citep{lyndenbell_stability_1967},
\[
    \left(\mathbf{L} - 2 m \omega \mathbf{R} + \omega^2 \bm{\Delta}\right)\mathbf{c} = 0,\label{eq:qhep1}
\]
solved separately for each \(m\) and \(\ell\). Here \(\mathbf{L}\) is the matrix representation of the nonrotating wave operator; in the natural basis of normal modes, it is diagonal, with elements \(- \omega_i^2 \delta_{n,n'}\). In that same basis, \(\bm{\Delta}\) is the identity matrix.

Neglecting the nonlinear near-degeneracy effects described in \citet{deheuvels_near_2017} and \obb, the rotating mode frequencies of \cref{eq:sym}, for modes of different \(m\), but the same \(n\) and \(\ell\), also arise from solving a conjugate Hermitian eigenvalue problem
\[
    \left(-\omega_{n\ell}^2 \mathbb{I}_{2\ell+1} - 2 \omega R_{n\ell} \mathbf{J}_z + \omega^2\mathbb{I}_{2\ell+1}\right)\mathbf{y} = 0 \label{eq:qhep2}
\]
whose solutions for each combination of \(n,m\) are the same as those from \cref{eq:qhep1} when \(\mathbf{R}\) is diagonal. Here \(\omega_{n\ell}\) are the nonrotating mode frequencies, and the components of the eigenvectors \(\mathbf{y}\) specify linear combinations of spherical harmonics, \(\sum_m y_m Y^\ell_m\), that constitute normal modes. By inspection, each of the matrices in \cref{eq:qhep2} is diagonal, and the normal-mode eigenvectors are the basis spherical harmonics, so this setup describes rotation around the \(z\)-axis of the reference coordinate system.

By azimuthal symmetry, we may set this coordinate system up so that the observer's line of sight lies in the \(xz\) plane. If the observer should in turn choose a different coordinate system, whose \(z'\)-axis is aligned with their line of sight, each of these normal modes may be written as linear combinations of spherical harmonics in the observer's coordinate system. The coefficients of these linear combinations are given simply by the columns of d-matrices \(\mathbf{d}^\ell(i)\), where \(i\) is the inclination angle between the \(z\)-axis and the line of sight. The relative observed amplitude of each eigenvector is given only by the norm of the \(m=0\) component of each normal mode in the observer's coordinate system \citep{gizon_inclination_2003}. It may thus be computed as
\[V = \mathbf{y}^\dagger \mathbf{d}^\ell(i)^\dagger \mathbf{P} \mathbf{d}^\ell(i) \mathbf{y},\label{eq:visintrinsic}\]
where the projection matrix
\[P_{m,m'} = \delta_{0,m}\delta_{0,m'}\label{eq:projection}\]
has only a single nonzero entry (on the diagonal at \(m=0\)).

Rather than being just spherical harmonics or just radial eigenfunctions, the actual nonrotating normal modes are products of both, when the unperturbed wave operator admits solutions under separation of variables. Accordingly, the full wave operator in general requires matrix elements to be taken with respect to \emph{products} of spherical harmonics and radial eigenfunctions. Correspondingly, the two quadratic Hermitian eigenvalue problems above, \cref{eq:qhep1,eq:qhep2}, are each submatrices of a uniquely defined, more general eigenvalue problem combining both of them. For a given \(\ell\), this is the tensor product of the two subproblems:
\[
\begin{aligned}
    &\left(\mathbf{L} \otimes \mathbb{I}_{2\ell + 1} - 2 \omega \mathbf{R}\otimes\mathbf{J}_z + \omega^2 \bm{\Delta} \otimes \mathbb{I}_{2\ell + 1}\right)(\mathbf{c} \otimes \mathbf{y})
    \\&\equiv\left(\tilde{\mathbf{L}} - 2 \omega \tilde{\mathbf{R}} + \omega^2\tilde{\bm{\Delta}}\right)\mathbf{x}
    \\&= 0.\label{eq:tensor1}
\end{aligned}
\]
Linear transformations \(\mathbf{A}\) and \(\mathbf{B}\) acting separately on \(n\) and \(m\), such that \(\xi_n \mapsto \sum_{n'} A_{n,n'} \xi_{n'}\) and \(Y^\ell_m \mapsto \sum_{m'} B_{m,m'} Y^\ell_{m'}\), may also be composed as \(\xi_n Y^\ell_{m} \mapsto \sum_{n'} \sum_{m'} A_{n,n'} B_{m,m'} \xi_{n'}Y^\ell_{m'} \equiv \sum_{m',n'} (\mathbf{A}\otimes \mathbf{B})_{nm, n'm'}\xi_{n'}Y^\ell_{m'}\). Thus constructed, the entries of the tensor-product matrices \(\tilde{\mathbf{L}}\) and \(\tilde{\mathbf{R}}\) are indexed by all possible combinations of \(n\) and \(m\), as are the eigenvectors emerging from the solution to \cref{eq:tensor1}. One may construct matrix representations possessing these elements using the Kronecker product, in which, for example, the scalar matrix elements \(L_{n,n'}\) of \(\mathbf{L}\) are each replaced by matrices \((\mathbf{\tilde{L}})_{n,n'} = L_{n,n'} \times \mathbb{I}_{2\ell+1}\), indexed by azimuthal order \(m\). In this same representation, the scalar matrix elements \cref{eq:rotkernel} of \cref{eq:qhep1} are now each replaced by integrals over matrices of the form
\[
     (\tilde{\mathbf{R}})_{n,n'}  = R_{n,n'} \times \mathbf{J}_z = B_{n,n'} \int \mathrm d r\ \Omega(r)\ \mathbf{J}_z\ K_{n,n'}(r).\label{eq:rotkernel2}
\]
\cref{eq:qhep1} can be seen to be recovered by taking the submatrices of \cref{eq:tensor1} under the Kronecker product with fixed \(m\), while \cref{eq:qhep2} is recovered with fixed \(n\).

\subsection{Radial Stratification of the Rotational Axis}\label{sec:misalignment}

Our preceding discussion has taken place in a coordinate system where the star's rotation determines the \(z\)-axis. However, there is no single preferred coordinate system if different mass shells should have different rotational axes. Thus, let us first consider these expressions as written in the observer's coordinate system, where the line of sight lies on the \(z'\) axis. As discussed in \autoref{sec:matrices}, rotating between coordinate systems transforms angular momentum matrices as \(\mathbf{J} \mapsto \mathbf{D}^\dagger\mathbf{J}\mathbf{D}\) for some Wigner D-matrix \(\mathbf{D}\), and in particular if \(\mathbf{e}_z \mapsto \hat{\mathbf{n}}\), then \(\mathbf{J}_z \mapsto \mathbf{D}^\dagger\mathbf{J}_z\mathbf{D} = \hat{\mathbf{n}} \cdot\vec{\mathbf{J}}\). The rotation from the star's to the observer's coordinate system is such that in the latter, the star's rotational axis points along \(\hat{\mathbf{n}}\). This being so, \cref{eq:rotkernel2} may be written in the observer's coordinate system as
\[
     (\tilde{\mathbf{R}})_{n,n'}  = R_{n,n'} \times (\hat{\mathbf{n}} \cdot\vec{\mathbf{J}}) = B_{n,n'} \int \mathrm d r\ \Omega(r)\ (\hat{\mathbf{n}} \cdot\vec{\mathbf{J}})\ K_{n,n'}(r).
\]
The expression for visibility factors using eigenvectors computed in the observer's coordinate system simplifies also from \cref{eq:visintrinsic} to
\[
    V = \mathbf{y}^\dagger \mathbf{P} \mathbf{y}.\label{eq:vis}
\]

Let us now relax the constraint of rotational alignment, and permit the rotational axis of each mass shell to be oriented independently of all the others. To describe this, we define a family of unit vectors \(\hat{\mathbf{n}}(r)\) indexed by radial coordinate, describing the orientation of the axis of rotation for the mass shell at that radius, so that the rotation rate of each mass shell is described by a vector \(\mathbf{\Omega}(r) = \Omega(r)\hat{\mathbf{n}}(r)\). By linearity --- i.e.~by \cref{eq:rotkernel} --- the total angular-momentum matrix elements are each the sum of contributions from each mass shell, which we may consider independently. In turn, each mass shell contributes only the integrand of \cref{eq:rotkernel2} evaluated at its radius, expressed in its own coordinate system. These contributions may each separately be transformed to the observer's line-of-sight coordinate frame by rotating against some unitary matrices \(\mathbf{D}^\ell(r)\): \(\mathbf{J}_z \mapsto {\mathbf{D}^\ell}^\dagger(r)\mathbf{J}_z\mathbf{D}^\ell(r) = \hat{\mathbf{n}}(r) \cdot\vec{\mathbf{J}}\), separately for each mass shell. Accordingly, the total angular momentum matrix has elements that sum over these contributions as
\[
\begin{aligned}
    (\tilde{\mathbf{R}})_{n,n'} &= B_{n,n'} \int \mathrm d r\ \Omega(r)\ ({\mathbf{D}^\ell}(r)^\dagger \mathbf{J}_z \mathbf{D}^\ell(r))\ K_{n,n'}(r)\\
    &= B_{n,n'} \int \mathrm d r\ \Omega(r)\ \left(\hat{\mathbf{n}}(r) \cdot \vec{\mathbf{J}}\right)\ K_{n,n'}(r)\\
    &= B_{n,n'} \left(\int \mathrm d r\ \mathbf{\Omega}(r)\ K_{n,n'}(r) \right) \cdot \vec{\mathbf{J}}\\
    &\equiv B_{n,n'}\ (\mathbf{\Omega}_{\text{eff},n,n'} \cdot \vec{\mathbf{J}}).\label{eq:rotkernel3}
\end{aligned}
\]
That is, to compute the matrix elements \((\tilde{\mathbf{R}})_{n,n'}\), one may equivalently first generate an effective angular momentum vector \(\mathbf{\Omega}_\text{eff}\) by classical vector addition --- as an average with respect to the rotational kernels \(K_{n,n'}\) --- and then map this vector to a spin matrix through \cref{eq:ndotJ}. We can express this more compactly by defining a matrix-valued function of the radial coordinate \(\mathbf{C}\) to have entries \(C_{n,n'}(r) = B_{n,n'}K_{n,n'}(r)\), so that
\[
    \tilde{\mathbf{R}} = \int \mathbf{C}(r) \otimes (\mathbf{\Omega}(r) \cdot \vec{\mathbf{J}})\ \mathrm d r.\label{eq:rotkernel4}
\]
A sum of tensor products may not itself necessarily be factorisable as a tensor product; we note this is now the case with \(\tilde{\mathbf{R}}\), as the direction of \(\mathbf{\Omega}_\text{eff}\) will in general not be the same for each radial-order index \(n\), or for the off-diagonal elements where \(n \ne n'\).

\section{A Two-zone Model}\label{sec:twozone}

\begin{figure}
\centering
\pandocbounded{\includegraphics[keepaspectratio]{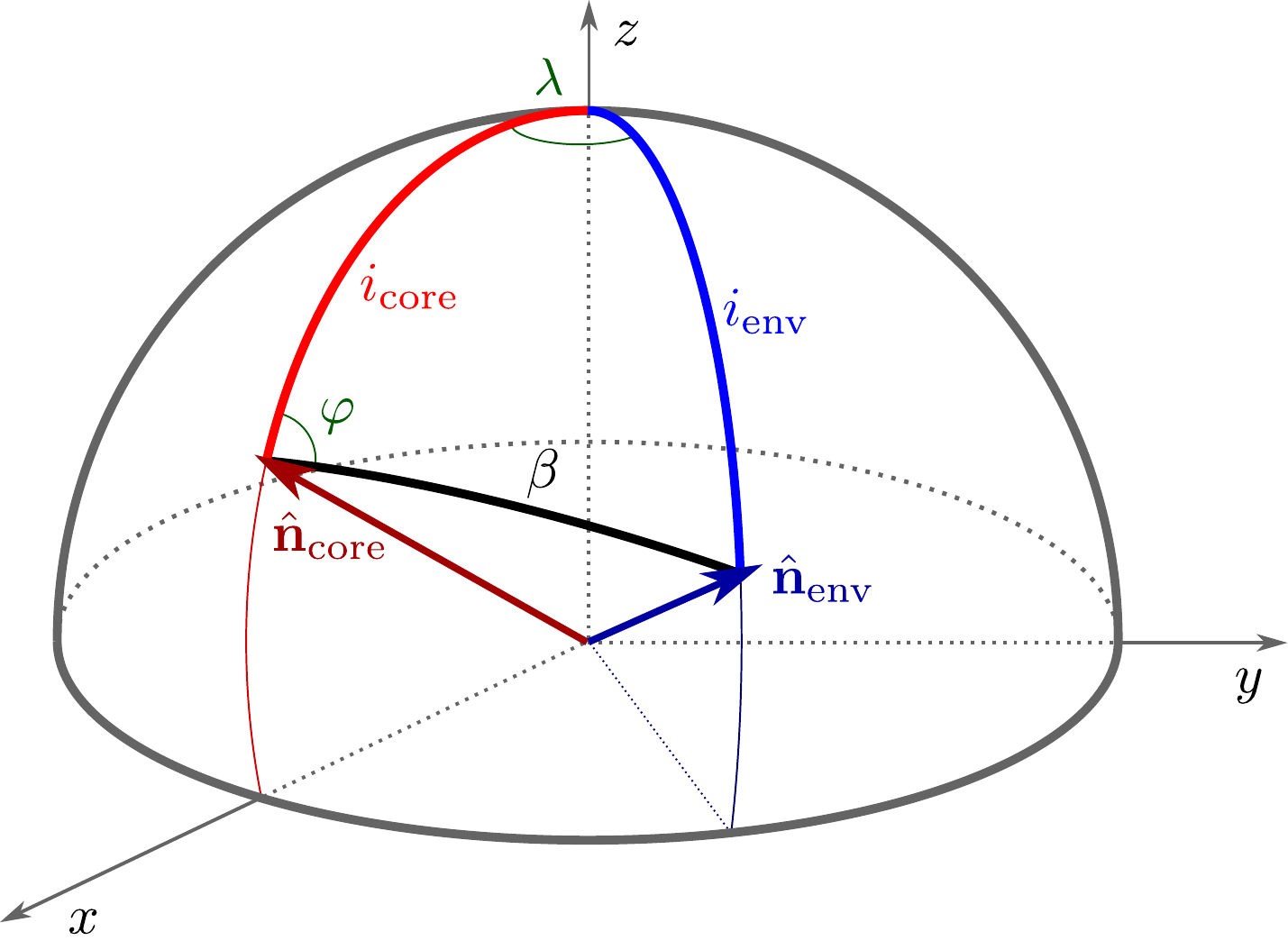}}
\caption{Parameterisation of core-envelope misalignment. The core and envelope are assumed separately to rotate around two axes specified by unit vectors \(\hat{\mathbf{n}}_\text{core}\) and \(\hat{\mathbf{n}}_\text{env}\), respectively. In the observer's coordinate frame, with the line of sight pointing downwards along the \(z\)-axis, these correspond to inclination angles \(i_\text{core}\) and \(i_\text{env}\). Only the misalignment angle \(\beta\) determines the frequency eigenvalues and rotational splittings, but it is underspecified by the two inclination angles. An additional obliquity angle \(\lambda\) needs also to be specified in order for the geometry of the configuration to be fully determined. Alternatively, the two-zone model can also be completely specified by the three angles \(i_\text{core}\), \(\beta\), and an orientation angle \(\varphi\).\label{fig:triangle}}
\end{figure}

To build intuition for this construction, we will apply it to a simplified two-zone model of radial differential rotation, of the kind originally proposed in \citet{gough_possible_1993}, wherein \(\mathbf{\Omega}\) is separately constant in the inner ``core'' and outer ``envelope''. The geometry of the problem in the observer's coordinate system (shown in \cref{fig:triangle}) is that of a spherical triangle, which may be fully determined either by way of two inclination angles \(i_\text{core}\) and \(i_\text{env}\) and an obliquity angle \(\lambda\), or by one of these inclination angles (say \(i_\text{core}\)), an intrinsic misalignment angle \(\beta\), and an orientation angle \(\varphi\). The rotational profile itself may be written as
\[
    \mathbf{\Omega}(r) = \left\{\begin{array}{cl}\mathbf{\Omega}_c = \Omega_c \hat{\mathbf{n}}_\text{core}, & r \le r_c \\ \mathbf{\Omega}_e = \Omega_e \hat{\mathbf{n}}_\text{env}, & r > r_c\end{array}\right..
\]
These two direction vectors are associated with two rotation matrices from the observer's coordinate system, and thus, through \cref{eq:ndotJ}, with angular momentum matrices:
\[
\begin{aligned}
\mathbf{D}^\ell_\text{core} = \mathbf{d}^\ell\left(i_\text{core}\right) \iff& \hat{\mathbf{n}}_\text{core}\cdot\vec{\mathbf{J}} = \cos i_\text{core}\mathbf{J}_z + \sin i_\text{core}\mathbf{J}_x\\
\mathbf{D}^\ell_\text{env} = \mathbf{d}^\ell\left(i_\text{env}\right) \exp\left[\ii \lambda \mathbf{J}_z\right] \iff& \hat{\mathbf{n}}_\text{env}\cdot\vec{\mathbf{J}} = \cos i_\text{env}\mathbf{J}_z \\
& + \sin i_\text{env} \cos \lambda \mathbf{J}_x\\
& + \sin i_\text{env} \sin \lambda \mathbf{J}_y.
\end{aligned}
\]

While \(i_\text{core}\) and \(i_\text{env}\) determine the visible mode amplitudes, the mode frequencies themselves are determined only by the internal misalignment angle \(\beta\), with
\[
    \cos\beta = \hat{\mathbf{n}}_\text{core} \cdot \hat{\mathbf{n}}_\text{env} = \cos i_\text{core} \cos i_\text{env} + \sin i_\text{core} \sin i_\text{env} \cos\lambda.
\]
This is because the eigenvalues of \cref{eq:tensor1} are determined only by
\[
\begin{aligned}
   \mathbf{D}_\text{env}\mathbf{D}_\text{core}^\dagger &= \mathbf{d}^{\ell}(i_\text{env}) \exp\left[\ii \lambda \mathbf{J}_z\right]\left(\mathbf{d}^{\ell}(i_\text{core})\right)^\dagger \\&= \exp\left[\ii \alpha \mathbf{J}_z\right] \mathbf{d}^{\ell}(\beta) \exp\left[\ii \gamma \mathbf{J}_z\right],
\end{aligned}
\]
which describes how one rotates from the core's coordinate system to that of the envelope. The other two Euler angles \(\alpha\) and \(\gamma\) specify rotations around either \(\hat{\textbf{n}}_\text{core}\) or \(\hat{\textbf{n}}_\text{env}\), without altering the relative misalignment of the two coordinate systems.

Let us first suppose that the interactions between multiplets may be ignored, permitting each multiplet only to be characterised by the angular momentum matrix \((\tilde{\mathbf{R}})_{nn}\) associated with that mode. The entries of \((\tilde{\mathbf{R}})_{nn}\) are specified entirely by the strictly positive on-diagonal rotational kernel, \(K_{nn}(r) \ge 0\), through \cref{eq:rotkernel3}. However, by the triangle inequality, \(\left|\int K(r) \mathbf{\Omega}(r) \mathrm d r \right| \le \int K(r) \ \left|\mathbf{\Omega}(r)\right| \mathrm d r\). In words, the first-order multiplet widths that would be obtained from a misaligned configuration are strictly less than or equal to that obtained from an otherwise identical configuration with all layers rotating around the same axis.

For each mode with index \(n\), we may define a core sensitivity parameter
\[
    a_n = \int_0^{r_c} K_{nn}(r) \mathrm d r,\label{eq:sensitivity}
\]
in terms of which \cref{eq:rotkernel3} gives
\[
    (\tilde{\mathbf{R}})_{nn} = B_{nn} \left(a_n \mathbf{\Omega}_c + (1-a_n) \mathbf{\Omega}_e\right) \cdot \vec{\mathbf{J}} \equiv B_{nn} (\mathbf{\Omega}_{\text{eff},n} \cdot \vec{\mathbf{J}}).
\]
Combining this with the corresponding entries of \(\tilde{\mathbf{L}}\) and \(\tilde{\bm{\Delta}}\), a la \cref{eq:qhep2}, gives
\[
    \left(-\omega_{n\ell}^2 \mathbb{I}_{2\ell+1} - 2 \omega B_{nn} (\mathbf{\Omega}_{\text{eff},n} \cdot \vec{\mathbf{J}}) + \omega^2\mathbb{I}_{2\ell+1}\right)\mathbf{y}_n = 0.
\]
The multiplet associated with that mode will have a width given by \(B_n |\mathbf{\Omega}_\text{eff}| \le B_n(a_n \Omega_c + (1-a_n)\Omega_e)\), and the apparent inclination implied by the distribution of power between its components will be set by the direction of \(\mathbf{\Omega}_\text{eff}\).

As the direction of \(\mathbf{\Omega}_\text{eff}\) varies from mode to mode in general, so too do the eigenvectors \(\mathbf{y}_n\). This means that different multiplets will, in addition to possessing different multiplet widths as already usually accounted for, possibly also exhibit different apparent inclination angles, lying between \(i_\text{core}\) and \(i_\text{env}\), in the presence of such internal misalignment. Should such variations in the apparent inclination angle be observed, this feature would be a sufficient condition for diagnosing the presence of internal misalignment. However, this is only a sufficient and not necessary condition: the apparent absence of such variations does not exclude the possibility of core-envelope misalignment, depending on the overall orientation of the misaligned configuration with respect to the observer's line of sight. In particular, as shown in \cref{fig:triangle}, the misalignment angle is only fully determined when an additional obliquity angle \(\lambda\) is also specified, supplementing the core and envelope inclination angles. This is analogous to the geometry of determining spin-orbit misalignments in transiting planets, where an obliquity angle \(\lambda_\text{pl}\) permits misaligment between stellar and orbital angular momenta even when both lie in the sky plane. Without any further inputs, this geometrical degeneracy renders the problem of constraining \(\beta\) observationally underdetermined from the mode amplitude ratios alone.

A two-zone model of differential rotation of this kind, without misalignment, is already in common use to describe rotation in evolved solar-like oscillators --- typically sub- and red giants --- wherein the ``core'' and ``envelope'' are separately probed by the g- and p- mode cavities of gravitoacoustic mixed modes that propagate in these stars. In the regime of high radial order (for both the p- and the g-mode cavities separately), the rotational splittings in these red giants may likewise be approximated with a linear combination of core and envelope rotation rates. These are conventionally written with respect to mixing fractions \(\zeta\), which take values between 0 (for pure p-modes) and 1 (for pure g-modes). The first-order scalar expression for the rotational splitting \citep{goupil_seismic_2013},
\[
    \delta\omega_{\text{mixed},nm} \sim  m\left[\zeta_n B_{g,n} \Omega_c + (1-\zeta_n) B_{p,n} \Omega_e\right], \label{eq:goupil}
\] then generalises in the presence of core-envelope misalignment to the angular-momentum matrix on the diagonal,
\[
    (\tilde{\mathbf{R}})_{\text{mixed}, nn} \sim (\zeta_n B_{g,n} \mathbf{\Omega}_\text{core} + (1-\zeta_n) B_{p,n} \mathbf{\Omega}_\text{env})\cdot \vec{\mathbf{J}} \equiv \delta\omega_{\text{eff}, n} (\hat{\mathbf{n}}_\text{eff} \cdot\vec{\mathbf{J}}).\label{eq:goupil2}
\]
We note that \(\zeta\) is related to the core sensitivity parameter of \cref{eq:sensitivity} as \[
    a = {B_g \zeta \over B_g \zeta + B_p (1-\zeta)}
\]
with \(B_p\to 1\) and \(B_g \to 1 - 1/\ell(\ell+1)\) in the asymptotic limit, so that for dipole modes, \(a_n \sim \zeta/(2-\zeta)\). Explicitly, in the observer's coordinate system, the linear splitting width and implied inclination angle for each multiplet become
\[
\begin{aligned}
\delta\omega_{\text{eff}, n}^2 =& \left(\zeta_n B_{g,n} \Omega_\text{core}\right)^2 + \left((1-\zeta_n) B_{p,n} \Omega_\text{env}\right)^2 \\ &+ 2\left(\zeta_n B_{g,n} \Omega_\text{core}\right)\left((1-\zeta_n) B_{p,n} \Omega_\text{env}\right)\cos\beta,\text{ and }\\
\cos i_{\text{eff}, n} =& \left(\zeta_n B_{g,n} \Omega_\text{core}\cos i_\text{core} + (1-\zeta_n) B_{p,n} \Omega_\text{env}\cos i_\text{env}\right) / \delta\omega_{\text{eff}, n},\label{eq:newgoupil}
\end{aligned}
\]
in this two-zone model of core-envelope misalignment. Thus,

\begin{enumerate}
\def\labelenumi{(\arabic{enumi})}
\tightlist
\item
  the multiplet splitting \(\delta\omega_\text{eff}\) will be smaller than the value of \(\delta\omega_\text{rot}\) returned from the standard first-order expression, \cref{eq:goupil}, as the magnitude of \(\mathbf{\Omega}_\text{eff}\) for each mode is reduced by misalignment. Departures from the first-order expression will in principle be largest for modes of intermediate character (i.e.~neither strictly p- nor g-dominated). Such modes in particular are, however, invariably found only when the underlying p- and g-modes come into resonance, in which case near-resonance effects arise which further modify the rotational splittings nonlinearly. We will discuss in more detail shortly.
\item
  the components of each mixed-mode multiplet will moreover exhibit relative visibilities that indicate intermediate inclination angles between \(i_\text{core}\) and \(i_\text{env}\), with the precise value being determined by the mixing fraction \(\zeta\) of that multiplet. These differences will, unlike point (1) above, be largest when comparing e.g.~the most p-dominated mixed modes against the most g-dominated ones.
\end{enumerate}

\subsection{Mixed Modes and Near-Degeneracy Effects}\label{mixed-modes-and-near-degeneracy-effects}

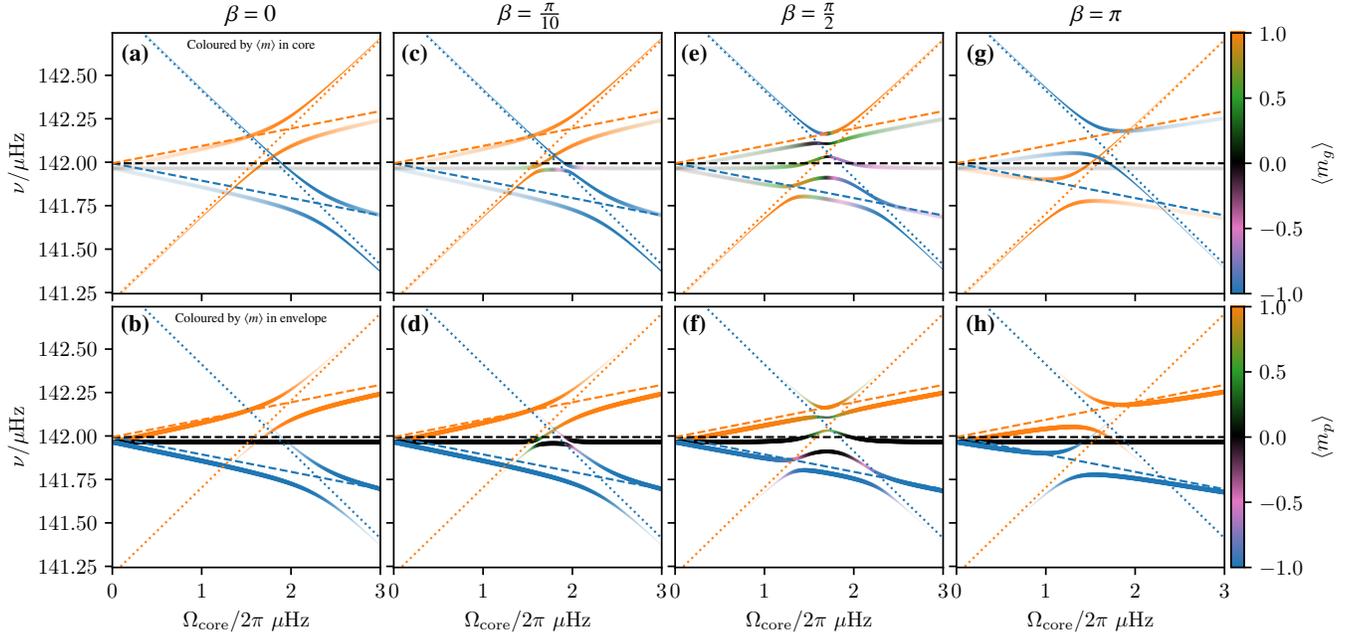
\begin{figure*}[htbp]
    \centering
    \annotate{\includegraphics[width=\textwidth, trim=0 .25cm 0 0.25cm, clip]{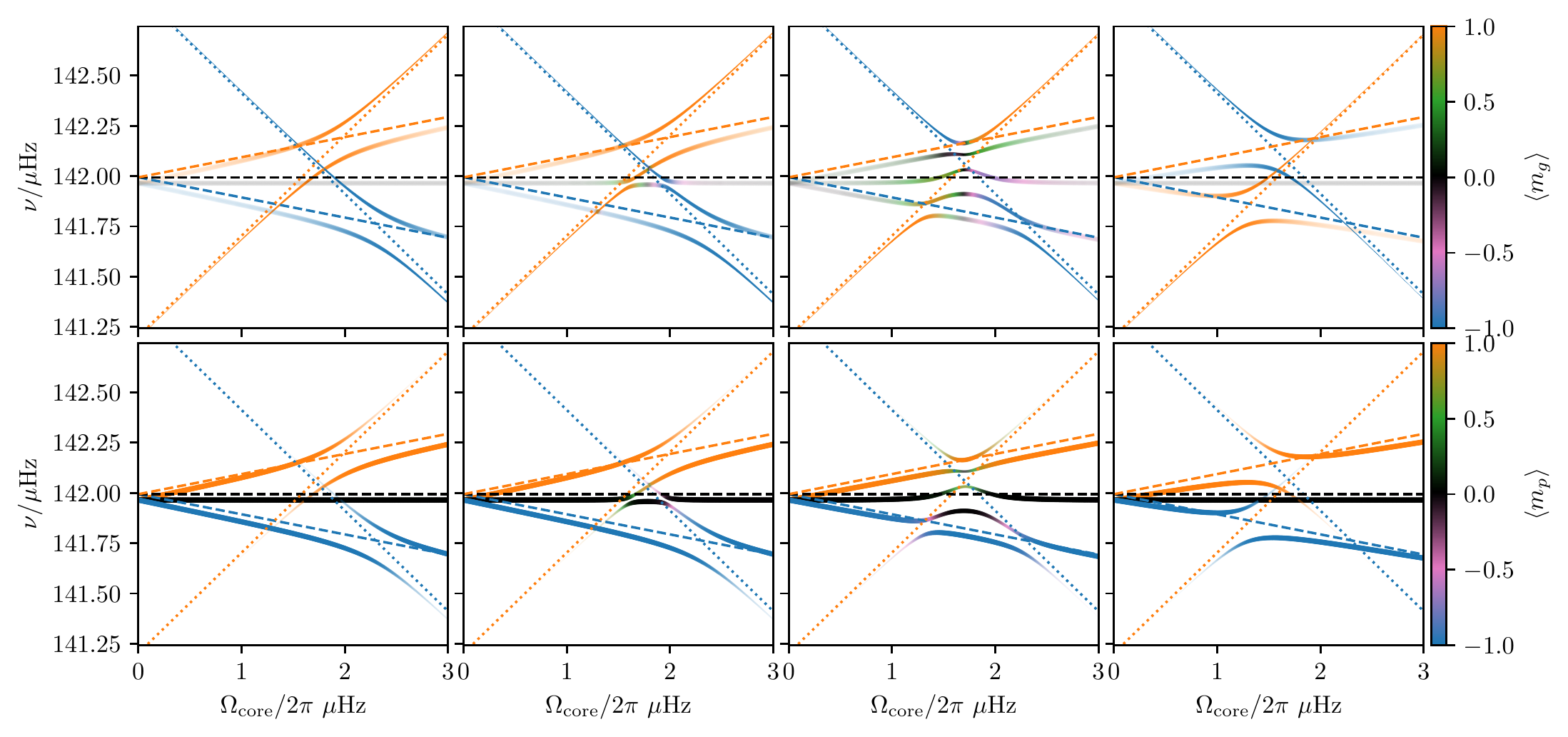}}{
    \node at (.19, 1.015){$\beta = 0$};
    \node at (.395, 1.015){$\beta = {\pi \over 10}$};
    \node at (.6, 1.015){$\beta = {\pi \over 2}$};
    \node at (.81, 1.015){$\beta = \pi$};
    \node at (.105, .95){\textbf{(a)}};
    \node at (.105, .505){\textbf{(b)}};
    \node at (.31, .95){\textbf{(c)}};
    \node at (.31, .505){\textbf{(d)}};
    \node at (.515, .95){\textbf{(e)}};
    \node at (.515, .505){\textbf{(f)}};
    \node at (.725, .95){\textbf{(g)}};
    \node at (.725, .505){\textbf{(h)}};
    \node at (.19, .965){\tiny Coloured by $\left<m\right>$ in core};
    \node at (.19, .515){\tiny Coloured by $\left<m\right>$ in envelope};
    }
    \caption{Modifications to avoided crossings in the presence of misalignment. All panels show rotational avoided crossings at a fixed core-envelope rotational contrast for a red-giant stellar model with artificially reduced coupling strength. Each column of panels shows a fixed value of the core-envelope misalignment angle $\beta$; the upper row of panels shows modes coloured by the effective azimuthal order $\left<m_g\right>$ of the mode around the core's rotational axis, while the lower row of panels shows modes coloured by the effective azimuthal order $\left<m_p\right>$ around the envelope's rotational axis. Dotted lines show the underlying rotationally-split pure g-mode frequencies, while dashed lines show the rotationally-split pure p-mode frequencies. The stroke width shows the extent to which a mode is p-dominated (i.e. the width is proportional to $1-\zeta$).}
    \label{fig:avoidedcrossings}
\end{figure*}

This first-order treatment of mixed-mode splittings does not account for the near-resonance effects which emerge in physical configurations where different multiplets interact with each other. Under such circumstances, \citet{deheuvels_near_2017} show that the off-diagonal elements of the matrix \(\mathbf{R}\) may no longer be ignored in the natural basis of mixed modes, which give rise to, e.g., different mixing fractions for different multiplet components associated with asymmetric rotational splittings. In principle, then, the integral expressions \cref{eq:rotkernel3} or \cref{eq:rotkernel4} will have to be evaluated for all possible combinations of \(n\) and \(n'\), rather than only on the diagonal entries, in order to correctly describe rotational splittings.

In the absence of azimuthal symmetry around a single rotational axis, it is not obvious that variable-separable normal modes, as would be returned from the usual perturbative approach, should be a good physical description of rotating mixed modes in a misaligned configuration. Nonetheless, we may in general still express these normal modes as linear combinations of variable-separable basis functions. Since the cores and envelopes of red giants are only weakly coupled pulsationally, one may use a notional nonorthogonal basis of isolated pure p- and g-modes to describe mixed modes, e.g.~as constructed using the \(\pi/\gamma\) isolation scheme of \citet[][hereafter \ob]{ong_semianalytic_2020}, in the aligned case.

We propose and now motivate the use of these basis functions in the misaligned case as well. We note that the cores and envelopes of red giants are known to mostly exhibit large contrasts in their rotation rates \citep[e.g.][]{li_asteroseismic_2024}, to such an extent that envelope rotation is often ignored altogether in the derivation of rotation rates from red-giant asteroseismology \citep[e.g.][\obb]{beck_fast_2012,gehan_core_2018,gehan_automated_2021,mosser_period_2018}. This implies relatively long timescales for rotational coupling, compared to evolutionary timescales on the red giant branch. Given that the core and envelope are both rotationally and pulsationally decoupled, the assumption that pure p- and g-mode eigenfunctions are separately amenable to separation of variables remains physically reasonable irrespective of whether the core and envelope are in or out of alignment. The nonorthogonal basis functions of \ob~are precisely such separately variable-separable solutions. In this sense, they form a more natural basis for misaligned calculations than the standard basis of normal modes. Thus, we will work with respect to this choice of basis in our subsequent discussion.

\obb~demonstrate that, in the aligned case, using this nonorthogonal basis has the desirable property of also, effectively, diagonalising \(\mathbf{R}\). This diagonalisation property is a consequence of the p-mode rotation kernels being essentially insensitive to the core, and the g-mode kernels to the envelope. Thus, in the misaligned two-zone model, the p- and g-mode angular momentum operators may be separately written, through \cref{eq:rotkernel3,eq:rotkernel4}, as tensor products of the diagonal matrices \(\mathbf{R}_\pi\) and \(\mathbf{R}_\gamma\), and the angular momentum matrices pointing in the respective directions of the core and the envelope; the overall matrix \(\tilde{\mathbf{R}}\) is, by linearity, the sum of the two. In this nonorthogonal basis, the matrices \(\mathbf{L}\) and \(\bm{\Delta}\) are also no longer diagonal, but they may nonetheless written in block matrix form as, e.g.,
\[
    \mathbf{L} = \begin{bmatrix}\mathbf{L}_{\pi\pi} & \mathbf{L}_{\pi\gamma} \\\mathbf{L}_{\pi\gamma}^\dagger & \mathbf{L}_{\gamma\gamma} \end{bmatrix},
\]
where the blocks lying on the diagonal are approximately diagonal, and those lying off the diagonal describe the coupling between the p- and g-mode subsystems.

Inserting these expressions into \cref{eq:tensor1} allows it to be written in block form as
\[
\begin{aligned}
\Bigg(\begin{bmatrix}
{\mathbf{L}_{\pi\pi}} & \mathbf{L}_{\pi\gamma} \\ \mathbf{L}_{\pi\gamma}^T & {\mathbf{L}_{\gamma\gamma}}
\end{bmatrix} \otimes \mathbb{I}_{2\ell+1}
 &+ 2 \omega \begin{bmatrix} {\mathbf{R}_\pi} \otimes {\mathbf{D}^\ell}_\text{env}^\dagger\mathbf{J}_z\mathbf{D}_\text{env}^\ell & 0 \\ 0 & {\mathbf{R}_\gamma} \otimes {\mathbf{D}^\ell}_\text{core}^\dagger\mathbf{J}_z\mathbf{D}_\text{core}^\ell \end{bmatrix} \\
 &+ \omega^2 \begin{bmatrix}
 \mathbb{I} & \bm{\Delta}_{\pi\gamma} \\ \bm{\Delta}_{\pi\gamma}^T & \mathbb{I}
 \end{bmatrix} \otimes \mathbb{I}_{2\ell+1}
 \Bigg)
\ \mathbf{x} = 0,
\end{aligned}\label{eq:tensor2}
\]
where \(\otimes\) again denotes the tensor product. In this tensor-product basis, the expressions for the mixing fractions of \ob~under unit normalisation (\(\mathbf{x^\dagger x} = 1\)) also generalise to
\[
    \zeta \sim \mathbf{x}^\dagger \left(\begin{bmatrix}0 & 0 \\ 0 & \mathbb{I}\end{bmatrix}\otimes \mathbb{I}_{2\ell + 1}\right) \mathbf{x},\label{eq:tensorzeta}
\]
while the overall mode visibility (combining projection effects and the modulation to the mode amplitude from mixing) may be written as
\[
    V \sim \mathbf{x}^\dagger \left(\begin{bmatrix}\mathbb{I} & 0 \\ 0 & 0\end{bmatrix}\otimes \mathbf{P}\right) \mathbf{x},\label{eq:tensorV}
\]
where \(\textbf{P}\) is the projection matrix, \cref{eq:projection}.

The indecomposability of \(\tilde{\mathbf{R}}\) into a tensor product of splitting and angular momentum matrices in the presence of misalignment qualitatively modifies the phenomenology of the near-resonance avoided crossings discussed in \citet{deheuvels_near_2017} and \obb. We illustrate these modifications qualitatively in \cref{fig:avoidedcrossings}, which shows rotational avoided crossings in the dipole modes of Model I from \obb, where the coupling strength (i.e.~off-diagonal matrices \(\mathbf{L}_{\pi\gamma}\) and \(\bm{\Delta}_{\pi\gamma}\)) has been scaled down by a factor of 10 to make the avoided crossings more visible. Since two different rotational axes are involved, we moreover make a distinction between two different, partial, estimators of the azimuthal order. In particular, since the azimuthal order of each mode is given in the aligned case by
\[
    m = \mathbf{x}^\dagger \left( \mathbb{I} \otimes (\hat{\mathbf{n}} \cdot \vec{\mathbf{J}})\right)\mathbf{x},
\]
the azimuthal orders of the pure p-modes, \(m_p\), and of the pure g-modes, \(m_g\), may be estimated with identical expressions when \(\hat{\mathbf{n}}\) points along the direction of the envelope or core rotation axis, respectively. By projecting into the p- and g-mode subspaces separately, we generalise this to estimate the effective azimuthal order of mixed modes around the core and envelope, respectively, as
\[
    \begin{aligned}
    \left<m_g\right> &= \mathbf{x}^\dagger \begin{bmatrix} 0 & 0 \\ 0 & \mathbb{I} \otimes \mathbf{D}_\text{core}^\dagger \mathbf{J}_z \mathbf{D}_\text{core}\end{bmatrix}\mathbf{x} / \zeta;\\
    \left<m_p\right> &= \mathbf{x}^\dagger \begin{bmatrix} \mathbb{I} \otimes \mathbf{D}_\text{env}^\dagger \mathbf{J}_z \mathbf{D}_\text{env} & 0 \\ 0 & 0\end{bmatrix}\mathbf{x} / (1-\zeta).
    \end{aligned}
\]
While the pure p- and g-modes (respectively, \(\zeta \to 0\) and \(\to 1\)) separately take integer values of \(m_p\) and \(m_g\), mixed modes may take noninteger values, as they are linear combinations of the two. In the upper row of panels in \cref{fig:avoidedcrossings}, we therefore colour modes by \(\left<m_g\right>\), and set their transparency to \(\zeta\), while in the lower row we colour them by \(\left<m_p\right>\), setting their transparency to \(1-\zeta\). Both \(m_p\) and \(m_g\) reduce to the usual \(m\) in the aligned case, \(\beta = 0\), shown in the leftmost column (panels a and b): modes of each \(m_p\) couple only to modes of the same \(m_g\), producing independent sets of avoided crossings for each \(m_p = m_g = m\). The morphology of these avoided crossings can however be seen to be change as the misalignment angle is increased, shown in the succeeding columns.

Qualitatively, small misalignment angles introduce coupling between modes of different \(m_p\) and \(m_g\). The coupling strength is given by the off-diagonal elements of the rotation matrix \(\mathbf{d}\), which nonetheless remains diagonally dominated --- we illustrate this in the second column (panels c and d) of \cref{fig:avoidedcrossings} for \(\beta = \pi/10\). For small misalignment angles \(\beta\), the coupling between between dipole modes of \(m_p=1\) and \(m_g=-1\) (and vice versa) is weaker than between \(m_p = 0\) and \(m_g = \pm1\) (and vice versa): from \cref{eq:dipoledmatrix}, the former goes as \(\sim \beta^2\), and the latter goes as \(\beta\). Thus, while the \(m_p=1\) and \(m_g=-1\) undergo avoided crossings with each other, these are too small to see on \cref{fig:avoidedcrossings}c,d.

As the misalignment angle increases, the coupling strength between pairs of p- and g-modes with \(m_p \ne m_g\) increases, while that between pairs with \(m_p = m_g\) decreases. At exact misalignment, with the core rotating at right angles to the envelope (\(\beta = \pi/2\)), modes of each \(m_p\) couple the most significantly to modes with \(m_g\) of opposite parity. For dipole modes in particular, we even have
\[
\small
    \mathbf{d}^{\ell = 1}\left({\pi \over 2}\right) = \begin{bmatrix}
 \frac{1}{2} & \frac{1}{\sqrt{2}} & \frac{1}{2} \\
 -\frac{1}{\sqrt{2}} & 0 & \frac{1}{\sqrt{2}} \\
 \frac{1}{2} & -\frac{1}{\sqrt{2}} & \frac{1}{2} \end{bmatrix},
\]
indicating that p-modes with \(m_p = 0\) only couple to g-modes of \(m_g = \pm 1\), and vice versa; the modified morphology of the resulting avoided crossings that this implies, which we illustrate in the third column (panels e and f) in \cref{fig:avoidedcrossings}, is qualitatively very different from the aligned case. Finally, reports in the literature of negative envelope rotation rates \citep[e.g.][]{deheuvels_seismic_2012, deheuvels_seismic_2014, triana_internal_2017} imply anti-aligned cores and envelopes (\(\beta = \pi\)). The avoided crossings of this configuration are illustrated in the rightmost column (panels g and h) in \cref{fig:avoidedcrossings}, with the sectoral p-modes coupling exclusively to g-modes of opposite azimuthal order.

\subsection{Implications for Existing Rotational Measurements}\label{sec:empirical}

It is currently common observational practice to analyse the mode frequencies and multiplet visibilities independently of each other, with the former used only to constrain rotation rates \citep[e.g.][]{gehan_core_2018, li_asteroseismic_2024}, and the latter only inclination angles \citep[e.g.][]{gehan_automated_2021, li_asteroseismic_2024}. However, these estimates of inclination returned from mode visibilities alone may not be sufficient to constrain the core-envelope misalignment angle \(\beta\), owing to the additional degree of freedom provided by the obliquity \(\lambda\). Moreover, since the mixed-mode coupling strengths, and therefore the mixing fractions \(\zeta\), can be seen to depend on the misalignment angle (given the above discussion), estimates of axial inclinations made when accounting for these near-resonance effects may in turn also differ substantially from those made without doing so. Conversely, these modified avoided crossings depend only on \(\beta\), and so constraining their features, by modelling near-resonance asymmetric splittings accounting for misalignment, may supplement the mode visibilities to break this geometrical degeneracy.

Since the rotational avoided crossings are modified by core-envelope misalignment, these modifications will interact with existing techniques for rotational characterisation that rely on modelling these avoided crossings, particularly when the mode frequencies alone are used for rotational measurement. To illustrate this, we will display rotating mixed-mode frequencies on diagnostic ``stretched'' period-echelle diagrams \citep[e.g.][]{mosser_period_2018, gehan_core_2018}, in which a coordinate transformation is applied to the mixed-mode frequencies that maps them to the nearest pure g-mode frequencies \citep{ong_rotation_2023}. Under this coordinate transformation, the ``stretched'' nonrotating g-mode frequencies are mapped to vertical columns on the period-echelle diagram, which separate into distinct families of smooth curves when only the pure g-modes are rotationally split.

\begin{figure}
\centering
\pandocbounded{\includegraphics[keepaspectratio]{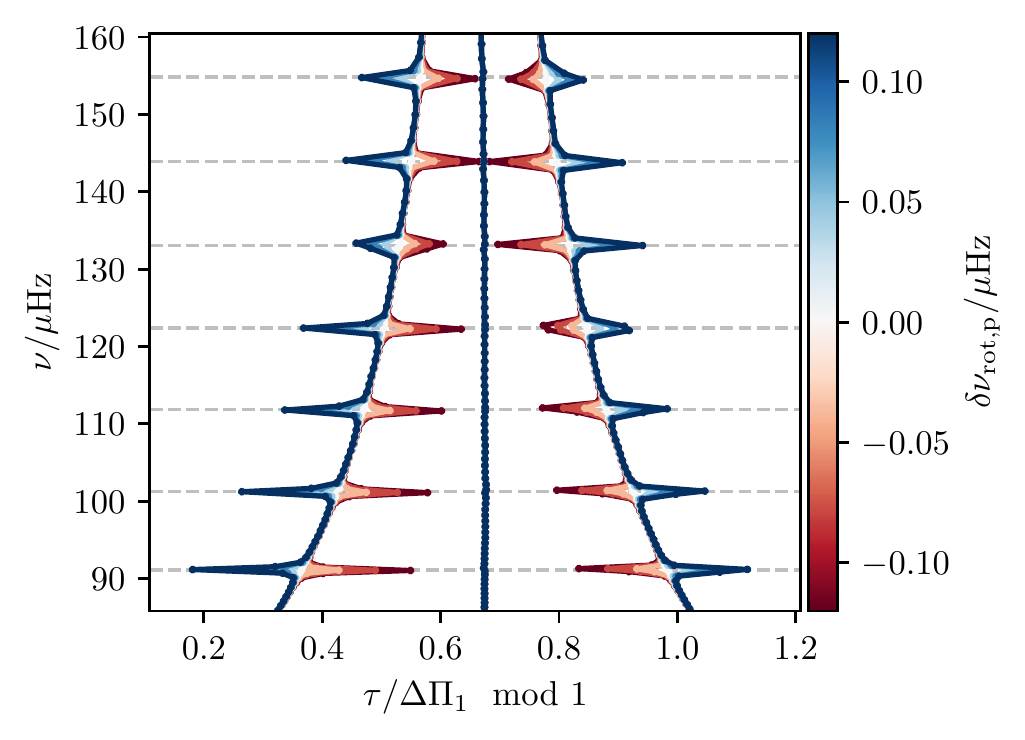}}
\caption{Stretched échelle diagrams for rotating mixed dipole modes, all generated with the same nonrotating stretching function, in the presence of envelope rotation. Each set of points shows the rotating mixed modes of a representative model of KIC 9267654, with a fixed amount of core rotation; they are coloured by the amount of envelope rotation included in the calculation. Points of the same \(m\) are joined with line segments of the same colour. Horizontal dashed lines show the locations of the underlying pure p-modes. Envelope rotation causes the most p-dominated sectoral modes to be shifted away from the zonal ones, in the opposite directions. They are perturbed away from the zonal modes for positive rotation rates.\label{fig:env}}
\end{figure}

The functional form of this coordinate transformation is derived from the same eigenvalue equation that gives rise to these avoided crossings in the asymptotic regime. As such, we may conversely use these diagrams to examine how existing techniques, relying on this description of avoided crossings, would respond to core-envelope misalignment. While a single stretching function does not yield uninterrupted rotationally-split ridges in the presence of envelope rotation \citep[\obb,][]{li_asteroseismic_2024}, we may still use the morphologies of the resulting curves for visualisation and qualitative discussion.

We illustrate how these diagnostic diagrams respond to envelope rotation in \cref{fig:env} in the purely aligned case. Specifically, we show a family of such stretched echelle diagrams, coloured by different values of the envelope rotation rate. These curves are generated by stretching the rotating mode frequencies of a \mesa~model constructed to match KIC 9267654 \citep{tayar_spinning_2022}, with other matrix elements evaluated using the expressions of \obb; for this series of curves, we also hold the core rotational splitting fixed at \(\delta\nu_\text{core} = 0.2\ \mu\)Hz. Note that we show both positive and negative values of the envelope rotation rate, as this existing formalism already accommodates envelope counterrotation.

Envelope rotation causes the most p-dominated rotating mode frequencies --- those closest to the pure p-mode frequencies, shown using horizontal dashed lines in \cref{fig:env} --- to be shifted away from the g-mode ridges on the stretched diagrams. The extent of this shift is determined by the p-mode rotation rate \(\delta\nu_\text{rot,p}\), and can be approximated by differentiating the analytic formula for the stretched period \(\tau\) of \citet{ong_rotation_2023} to yield
\[
     \delta\tau \sim {\partial \tau \over \partial \nu_p} \cdot m \delta\nu_\text{rot,p} = -m \left({1 \over \zeta} - 1\right){\delta\nu_\text{rot,p} \over \nu^2};
\]
we recall that \(1/\zeta - 1\) is well-approximated as a series of Lorentzians with widths \(q \Delta\nu/\pi\) and heights \(\nu^2 \Delta\Pi_\ell / q\Delta\nu\), centered at each p-mode frequency.

\begin{figure}[htbp]
    \centering
    \includegraphics[width=.475\textwidth,trim=0 .25cm 0 .25cm, clip]{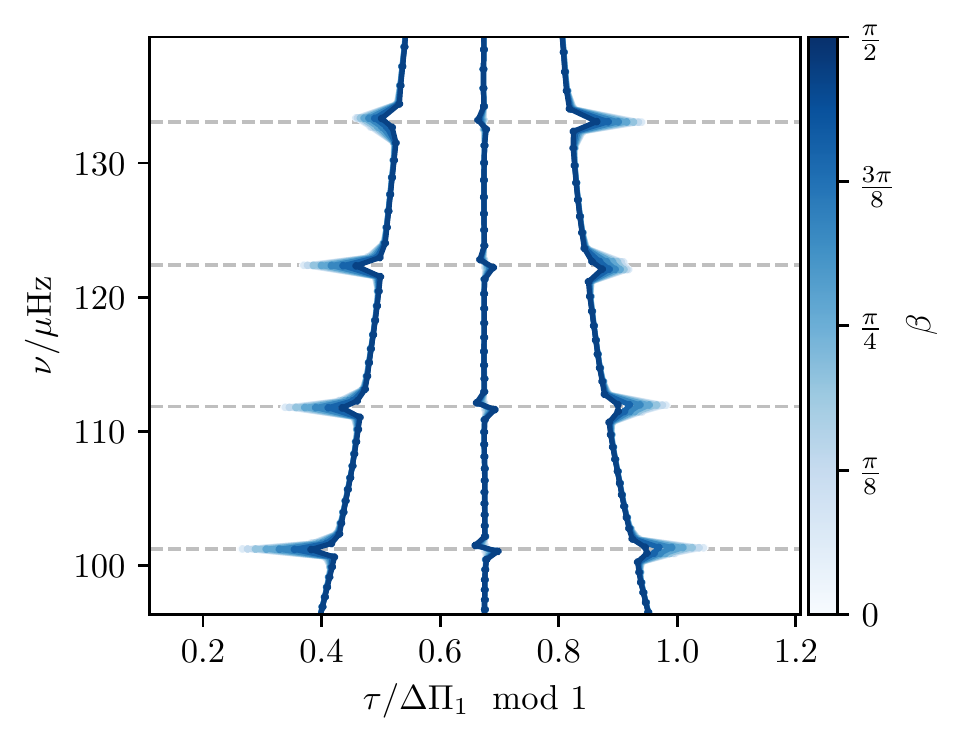}
    \includegraphics[width=.475\textwidth,trim=0 .25cm 0 .25cm, clip]{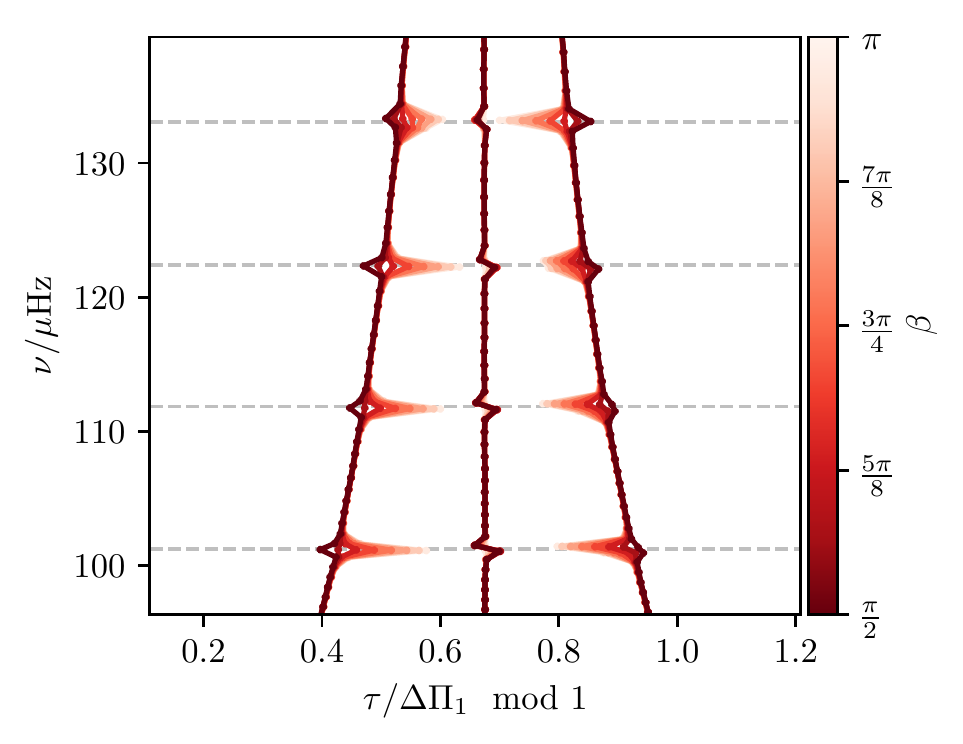}
    \caption{Modified morphology of the stretched period-echelle diagram in the presence of core-envelope misalignment. Rotating mode frequencies are shown with points as in \cref{fig:env}, and are coloured by the misalignment angle $\beta$. Modes which would have the same $m$ as $\beta \to 0$ are joined with line segments. To show the morphology more clearly, the upper panel shows the rotationally-split ridges for $0\le\beta\le{\pi\over2}$, while the lower panel shows them for ${\pi\over2}\le\beta\le\pi$. Horizontal lines show the locations of the underlying pure p-modes.}
    \label{fig:beta}
\end{figure}

In \cref{fig:beta}, we now illustrate how the morphology of this diagram is modified when the core and envelope come out of alignment. Holding the envelope rotation rate fixed at \(0.12\ \mu\)Hz (the largest value shown in \cref{fig:env}), the coloured curves of \cref{fig:beta} indicate the stretched-echelle power diagrams that would emerge for different values of the misalignment angle \(\beta\). For clarity, we show only positive values of \(\beta\) in the upper panel, and only negative values below. As the value of \(\beta\) increases, the deviations of the most p-dominated mixed modes away from the rotationally-split g-mode ridges can be seen to decrease.

These changes are difficult to distinguish from those associated with simply a reduced envelope rotation rate, as shown in \cref{fig:env}. While the morphology of the central ridge, consisting of zonal (\(m = 0\)) modes, can also be seen to be slightly modified by misalignment, an accounting of these modes does not ordinarily enter into existing measurements of rotation rates made using mixed modes; in any case, considerably better observational precision is required to detect these smaller changes, compared to that required for rotational measurement per se. Qualitatively, therefore, this discussion suggests that existing rotational measurement techniques, including those relying on the asymptotic procedure, will underreport the magnitude of the envelope rotation rate when confronted with an internally misaligned system.

\section{Case Study: Kepler-56}\label{sec:k56}

We will now apply this new theoretical formalism to reanalyse the asteroseismic planet host Kepler-56, which, having been extensively characterised, is known to be unique in many respects. \citet{huber_stellar_2013} find that it is rotationally anomalous: from photometric variability attributable to spot modulations, it is known to possess a surface rotational period of \(74 \pm 3 \,\mathrm{d}\), which would a priori imply a p-mode rotational splitting of \(0.156 \pm 0.006\,\mu\mathrm{Hz}\) --- considerably faster than would be predicted by existing single-star models of post-main-sequence angular-momentum transport and magnetic braking. Moreover, it is a particularly high signal-to-noise ratio example of a star on the lower RGB exhibiting rotationally-split gravitoacoustic mixed modes, potentially permitting core and envelope rotation to be disentangled. Most saliently to this work, the estimated inclination of its rotational axis was measured in \citet{huber_stellar_2013} to be intermediate between being pole-on and equator-on with high statistical confidence. This last point was particularly notable given that Kepler-56 hosts multiple transiting planets; it remains one of only a handful of multi-transiting planetary systems where a rotational axis of the host is out of alignment with multiple planets, which are themselves mutually aligned \citep[e.g.][]{ahlers_spinorbit_2015, hjorth_backward_2021, zhang_long_2021}.

\begin{figure}[htbp]
    \centering
    \includegraphics[trim=0 0.5cm 0 0.25cm, width=.45\textwidth]{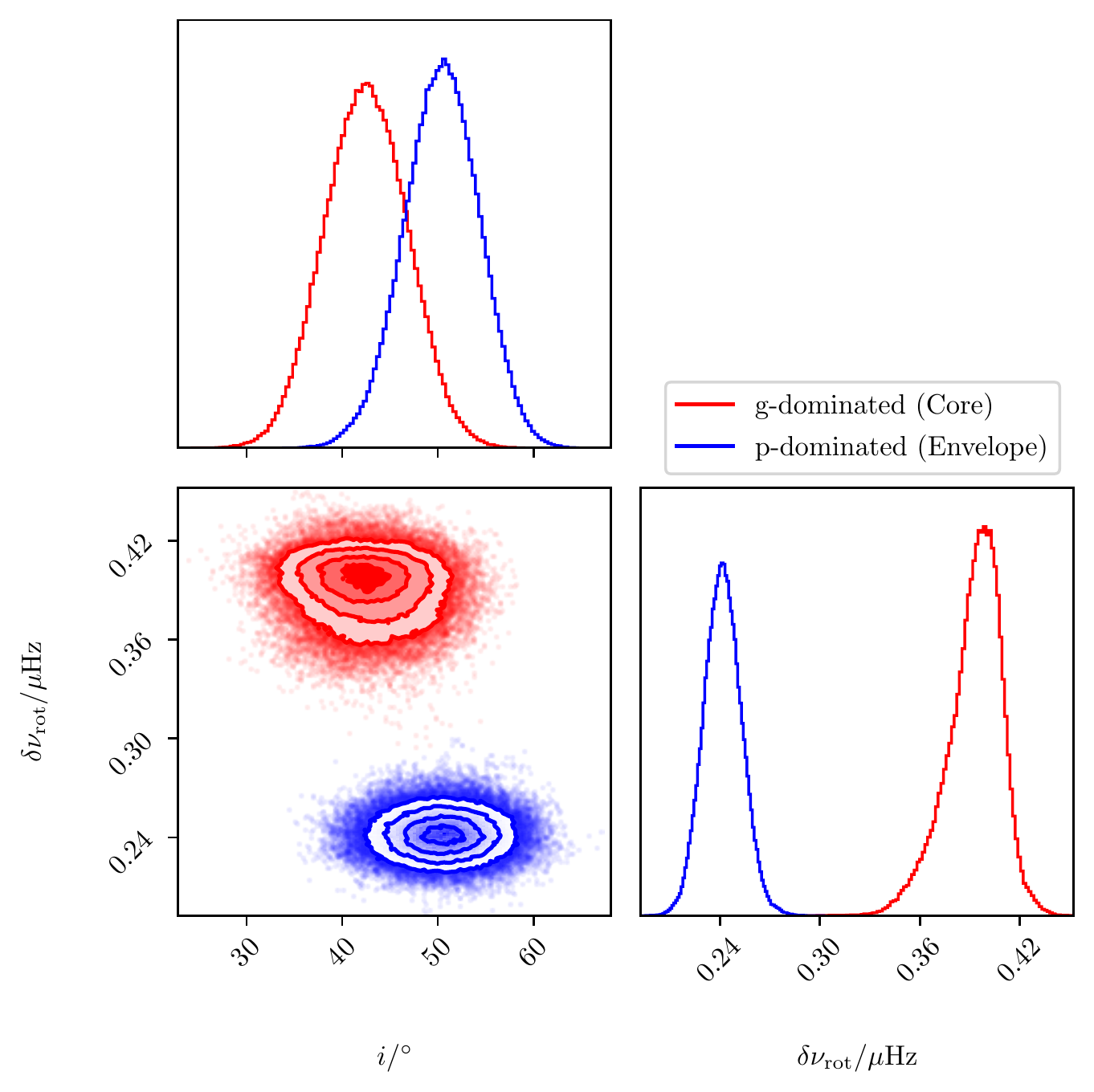}
    \caption{Summary of rotational characterisation of Kepler-56 in \cite{huber_stellar_2013}. The joint posterior distributions of the apparent inclination angle $i$ and the dipole-mode rotational multiplet widths $\delta\nu_\text{rot}$, were estimated in that work by averaging over the three most g-dominated (red) and p-dominated (blue) mixed modes, each fitted separately.}
    \label{fig:huberk56orig}
\end{figure}

In \citet{huber_stellar_2013}, this rotational characterisation was restricted to the 3 most p- and g-dominated multiplets, which were separately analysed to derive inclination and rotation-rate estimates. This invites reanalysis for several reasons. For one, since this procedure predates the publication of the first-order expression \cref{eq:goupil} \citep{goupil_seismic_2013}, no attempt was made to determine, or correct for, the effect of the mixing fraction \(\zeta\). However, while the most g-dominated modes in red giants are representative of the pure g-modes, even the most p-dominated modes exhibit significant sensitivity to the core \citep[e.g.][]{ahlborn_improved_2022, ong_redgiant_2024}; for example, the most p-dominated multiplets for Kepler-56 have mixing fractions of \(\zeta \sim 0.3\).

Given these limitations of their procedure, \cref{eq:goupil2} indicates that we should therefore expect the reported axial inclination angles of \citet{huber_stellar_2013}, for both the p- and g-dominated modes, to have been essentially dominated by core rotation: this is suggested by their inferred envelope rotation rate being much higher than the photometric surface rotation rate (bottom right panel of \cref{fig:huberk56orig}). However, even with their reduced specific sensitivity to the envelope, the posterior distribution for the inclination angle from p-dominated modes in \citet{huber_stellar_2013} is nontrivially different from that obtained via the g-dominated modes (top left panel of \cref{fig:huberk56orig}). This therefore suggests that the true posterior distribution for the envelope inclination angle, even only naively correcting for the mixing fraction, would have likely been considerably different from that for the core inclination angle, which would in turn imply significant misalignment between the core and envelope.

As such, while the seismic estimates of core rotation (both inclination and rate) in \citet{huber_stellar_2013} are likely to be accurate, those reported of the envelope were almost certainly heavily contaminated by sensitivity to the core. Thus, even setting aside the question of internal misalignment, a reanalysis of Kepler-56 would have been necessary anyway, if only to more accurately characterise its seismic envelope rotation rate. We pursue this reanalysis in this section, explaining our methodology accommodating misalignment in \autoref{sec:k56analysis}, describing our results in \autoref{sec:k56results}, and discussing their implications in \autoref{sec:k56discussion}.

\subsection{Revised Asteroseismic Rotational Analysis}\label{sec:k56analysis}

Following the above discussion, placing constraints on misalignment relies on accurate characterisation of both mode frequencies and (relative) amplitudes. Given the stochastic nature of solar-like oscillations, mode amplitudes are more accurately estimated from longer time series. Thus, for much the same reasons as \citet{huber_stellar_2013}, we opt to reanalyse the full four-year time series, available at long cadence, rather than the short-cadence time series, which is only available for two years. For this purpose, we use the \textsc{kepseismic} power-spectrum data products \citep{kepseismic1, kepseismic2, kepseismic3}, which are optimised for asteroseismology.

Before examining misalignment, we first re-characterised solar-like oscillations in the full Kepler time series using existing automated tools as far as possible. In particular, we fitted a model of the power-spectral density (PSD) of the radial and quadrupole modes against a background-divided power spectrum using \texttt{PBJam} \citep{pbjam}. We then fitted initial guesses at the asymptotic dipole-mode parameters --- including the dipole p-mode small separation \(\delta\nu_{01}\), g-mode period spacing \(\Delta\Pi_1\) and phase offset \(\epsilon_g\), core and envelope rotational splittings \(\delta\nu_\text{rot,g}\) and \(\delta\nu_\text{rot,p}\), an averaged rotational inclination angle \(i_\star\), and the strength of the coupling between the two mode cavities --- using \texttt{reggae} \citep{ong_reggae_2024}. These initial fits were used only to specify prior distributions for our subsequent analysis.

Placing constraints on internal misalignment will require us to combine constraints on mode frequencies and multiplet visibilities. These are conventionally first fitted from the power spectrum and then interpreted separately \citep[e.g.][]{gehan_core_2018, gehan_automated_2021, hall_weakened_2021, li_asteroseismic_2024}. Since no prescription exists by which to incorporate these separate fits into a combined analysis, we instead derive our constraints on misalignment directly from fitting a model of the power spectrum, which accepts multiple rotation rates and inclination angles as parameters to be inferred. These parameters are fitted by characterising discrepancies between this model and the power spectrum using the standard \(\chi^2\) likelihood function with two degrees of freedom.

Only the dipole modes of Kepler-56 exhibit mixing between the p- and g-mode cavities. As in \texttt{reggae}, to isolate this analysis to dipole modes only, we use a version of the power spectrum where the granulation background, and the model of the \(\ell = 0,2\) modes from \texttt{PBJam}, have been divided out. Notably, \citet{huber_stellar_2013} report the detection of octupole (\(\ell = 3\)) modes, which we do not explicitly include in our analysis. However, we obtain identical results whether or not we mask out regions of the power spectrum containing octupole modes when fitting a PSD model.

We adopt a PSD model which is largely identical to that of \texttt{reggae}. The mode visibilities \(V\), linewidths \(\Gamma\), and p-mode amplitudes \(H\) (which are assumed to vary according to a Gaussian envelope around \(\nu_\text{max}\), inherited from the initial fit from \texttt{PBJam}, and whose overall height \(H(\numax)\) is fitted as an additional free parameter), and rotating mixed-mode frequencies \(\nu_i\) are used to generate a sum of Lorentzians
\[
     \mathcal{P}(\nu) = 1 + \sum_i^{N_\text{modes}} {H(\nu_i) V_i \over 1 + 4{\left(\nu - \nu_i\right)^2 / \Gamma_i^2}}.
\]
In order to constrain internal misalignment from this fitting procedure, we extend this PSD model from that of \texttt{reggae} in the following respects:

\begin{enumerate}
\def\labelenumi{\arabic{enumi}.}
\tightlist
\item
  We permit the dipole-mode small separation to vary independently for each radial order, to accomodate small deviations arising from the dipole modes having sampled acoustic glitches at different phases from the radial and quadrupole modes \citep[e.g.][]{dreau_dipolar_2020, saunders_evolutionary_2023}, rather than using a single value for all radial orders.
\item
  We generalise the parameterisation of rotational orientation to include the three angles \(i_\text{core}\), \(i_\text{env}\), and \(\lambda\) shown in \cref{fig:triangle}, rather than a single inclination angle \(i_\star\). To accommodate this change, the mode coupling calculations are now performed for all \(m\) at once using \cref{eq:tensor2}, rather than separately for each \(m\) using \cref{eq:qhep1}. Mixing fractions and visibilities are then assigned to each mode using the generalised expressions \cref{eq:tensorzeta,eq:tensorV}.
\item
  Rather than use a fixed, artificially inflated, linewidth for all modes as in \texttt{reggae}, we now also modulate the linewidths by a factor of \(1-\zeta\), following \citet{benomar_evolved_2014}, so that g-dominated modes are modelled with narrower lines. The factor of \(1-\zeta\) applied to mode heights in the PSD model, which ultimately originates from the modes also having reduced amplitude \citep{belkacem_angular_2015}, is only applied when the mode linewidths fall below a critical resolution threshhold, resulting in dilution \citep{dupret_theoretical_2009}. However, to avoid the degeneracies that are potentially introduced when linewidths and rotational splittings are separately permitted to vary freely \citep[e.g.][]{kamiaka_reliability_2018}, we still hold the maximum (p-dominated) linewidth fixed at the average radial-mode values reported in \citet{huber_stellar_2013}.
\item
  Our best-fitting value of the g-mode phase offset \(\epsilon_g\) returned from \texttt{reggae} was significantly higher than the asymptotic value of \(\sim0.8\). While the g-mode rotational multiplets of Kepler-56 are symmetrically split, anomalous measurements of \(\epsilon_g\) may suggest the presence of weak magnetic fields in the core \citep{deheuvels_strong_2023}, which would also yield additional magnetic damping for g-modes that is not accounted for in the usual scaling of the linewidth with mixing fraction. Therefore, as in \texttt{reggae}, we apply some artificial broadening to the model of the power spectrum, by using a larger value for the minimum g-mode linewidth than the Rayleigh resolution limit of the power spectrum. Since constraints on magnetic damping are not the focus of this work, we hold this value fixed rather than varying it as a free parameter. We found that a threshold of \(\Delta\nu/400\), which is roughly ten times the usual resolution-limited value, gave reasonable agreement with the g-dominated modes in the power spectrum. This has the effect of smoothing out the likelihood-function landscape for the mixed-mode parameters \(\Delta\Pi\), \(\epsilon_g\), \(\alpha_g\), and the coupling strengths, which would otherwise yield sharp local maxima. Although this improves the numerical stability of our inference procedure, it also renders our posterior distributions broader than without such smoothing.
\end{enumerate}

\begin{figure}
\centering
\pandocbounded{\includegraphics[keepaspectratio]{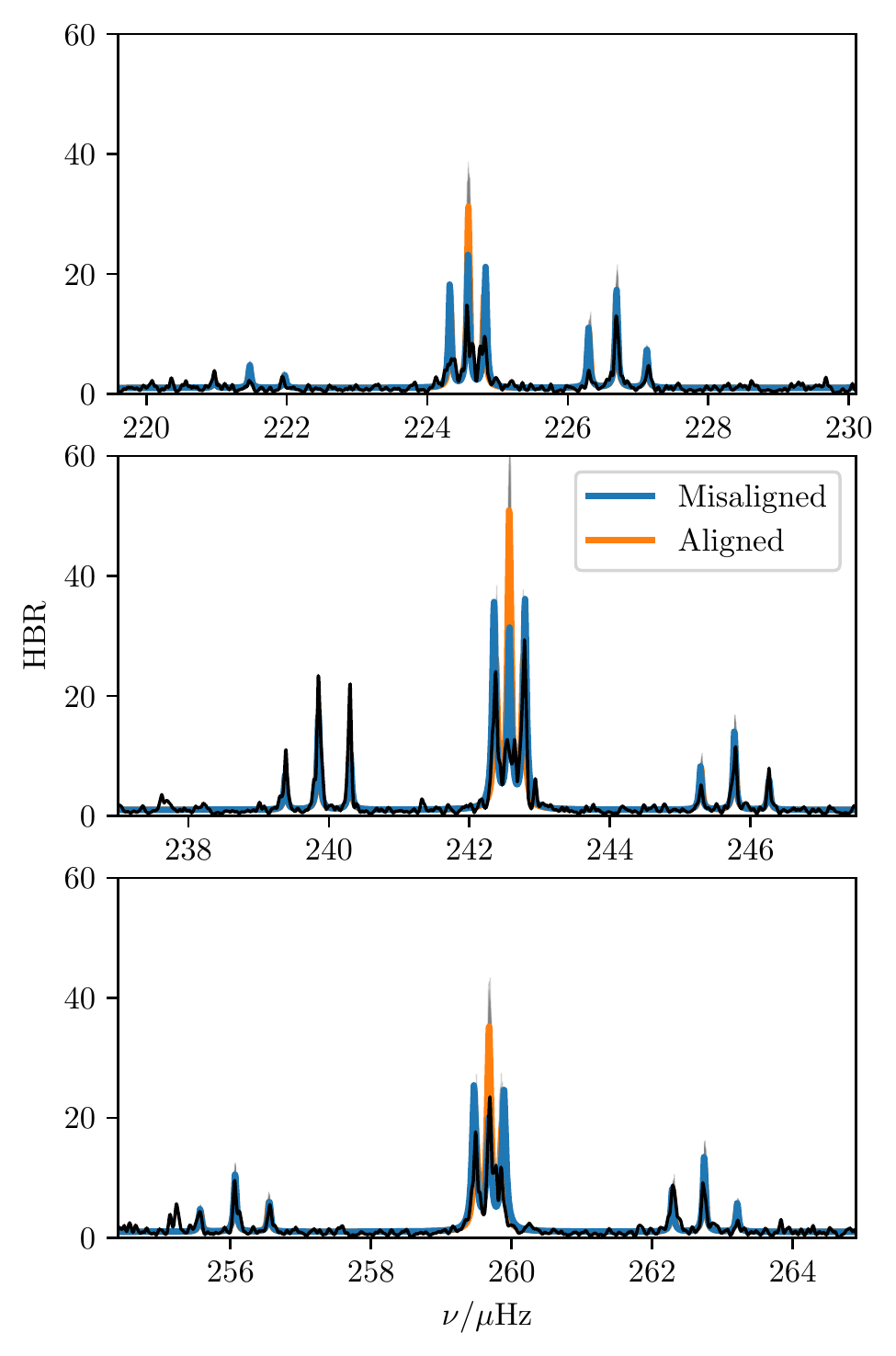}}
\caption{Fits for for a model of the power spectrum incorporating misaligned rotational mode-coupling calculations, overplotted against the 4-year Kepler power spectrum of Kepler-56 (black, smoothed by a Gaussian filter with width \(\sigma\) of 2 resolution elements), where the \(\ell = 0,2\) modes, and the overall granulation background, have both been divided out; the vertical axis therefore shows the height-to-background ratio, labelled HBR. Each panel shows half an echelle order, centered on the notional dipole p-mode frequency. The middle panel in particular shows the multiplets around the dipole p-mode closest in frequency to \(\numax\), whose relative mode heights differ significantly: the p-dominated one in the centre appears consistent with a higher projected inclination than the two, more g-dominated, multiplets on either side. The misaligned model (blue) is better able to accommodate this morphology than one with only a single rotational axis (orange). Pale gray curves in the background indicate draws from the posterior distribution, to illustrate statistical uncertainties in constraining the model.\label{fig:ps}}
\end{figure}

\subsection{Results}\label{sec:k56results}

\begin{table}[htbp]
\caption{Prior distributions on parameters of PSD model\label{tab:prior}}
\vspace*{-2em}
\begin{equation*}
\begin{array}{rl}
\text{p-mode parameters} &
    \left\{
    \begin{array}{rcl}
    \hspace{3.0em}\delta\nu_{01}(n)/\Delta\nu & \sim & \mathcal{U}(0.01, 0.05) \\
    \end{array}
    \right.\\
\text{g-mode parameters} &
    \left\{
    \begin{array}{rcl}
    \hspace{4.6em}\Delta\Pi_1 / \mathrm{s} & \sim & \mathcal{U}\left(86.62, 87.50\right) \\
    \epsilon_g & \sim & \mathcal{P}\left(0.5, 1.5\right) \\
    \alpha_g & \sim & \mathcal{U}\left(-4\times10^{-5}, 1.6\times10^{-4}\right)
    \end{array}
    \right.\\
\text{p/g mode coupling} &
    \left\{
    \begin{array}{rcl}
    \hspace{6.1em}p_L & \sim & \mathcal{U}(-2, 2) \\
    p_D & \sim & \mathcal{U}(15, 28)
    \end{array}
    \right.\\
\text{Rotation} &
    \left\{
    \begin{array}{rcl}
    \log_{10}\left(\delta\omega_\text{rot,g}/\mathrm{Hz}\right) & \sim & \mathcal{U}(-11, -5) \\
    \log_{10}\left(\delta\omega_\text{rot,p}/\mathrm{Hz}\right) & \sim & \mathcal{U}(-11, -5) \\
    \cos i_\text{core} & \sim & \mathcal{R}(0, 1) \\
    \varphi & \sim & \mathcal{R}(0, \pi) \\
    \beta & \sim & \mathcal{U}(0, \pi) \\
    \end{array}
    \right.\\
\text{Normalisation} &
    \left\{
    \begin{array}{rcl}
    \hspace{1.75em}\log_{10} \mathrm{H}\left(\numax\right) & \sim & \mathcal{U}\left(\log_{10}40, \log_{10}55\right) \\
    \end{array}
    \right.
\end{array}
\end{equation*}
\footnotesize \textsc{Notes} --- Here $\mathcal{U}(a, b)$ refers to the uniform distribution with support on the open interval $(a, b)$; $\mathcal{P}(a, b)$ refers to the uniform distribution on the interval $[a, b)$, where the endpoints are topologically identified under periodic boundary conditions; and $\mathcal{R}(a, b)$ refers to the uniform distribution on the interval $[a, b]$, sampled under reflective boundary conditions. $\alpha_g$ is a g-mode curvature parameter, while $p_L$ and $p_D$ are parameters describing the coupling matrices $\mathbf{L}$ and $\bm{\Delta}$, scaled in \texttt{reggae} with respect to values computed from a reference red-giant \mesa\ model. Other parameters are as described in the text. Five different values of $\delta\nu_{01}$ are fitted: one for each radial order.
\end{table}

To illustrate the need for this misaligned model, we show in the central panel of \cref{fig:ps} the multiplets closest in frequency to \(\numax\). The amplitude of the \(m=0\) peaks in the most p-dominated multiplets suggested by the power spectrum (shown in gray), relative to the \(m = \pm 1\) components, appear lower compared to adjacent, more g-dominated multiplets. This is suggestive of a large misalignment angle between the core and envelope, with the envelope seen closer to equator-on than the core. To guide the eye, we show the best-fitting (maximum-likelihood) model in the aligned scenario in orange, and a misaligned model in blue. While they produce identical predictions for the more g-dominated multiplets, the misaligned model can be seen to clearly better describe the p-dominated multiplet than the aligned one.

\begin{figure*}[htbp]
    \centering
    \annotate{\includegraphics[width=\textwidth]{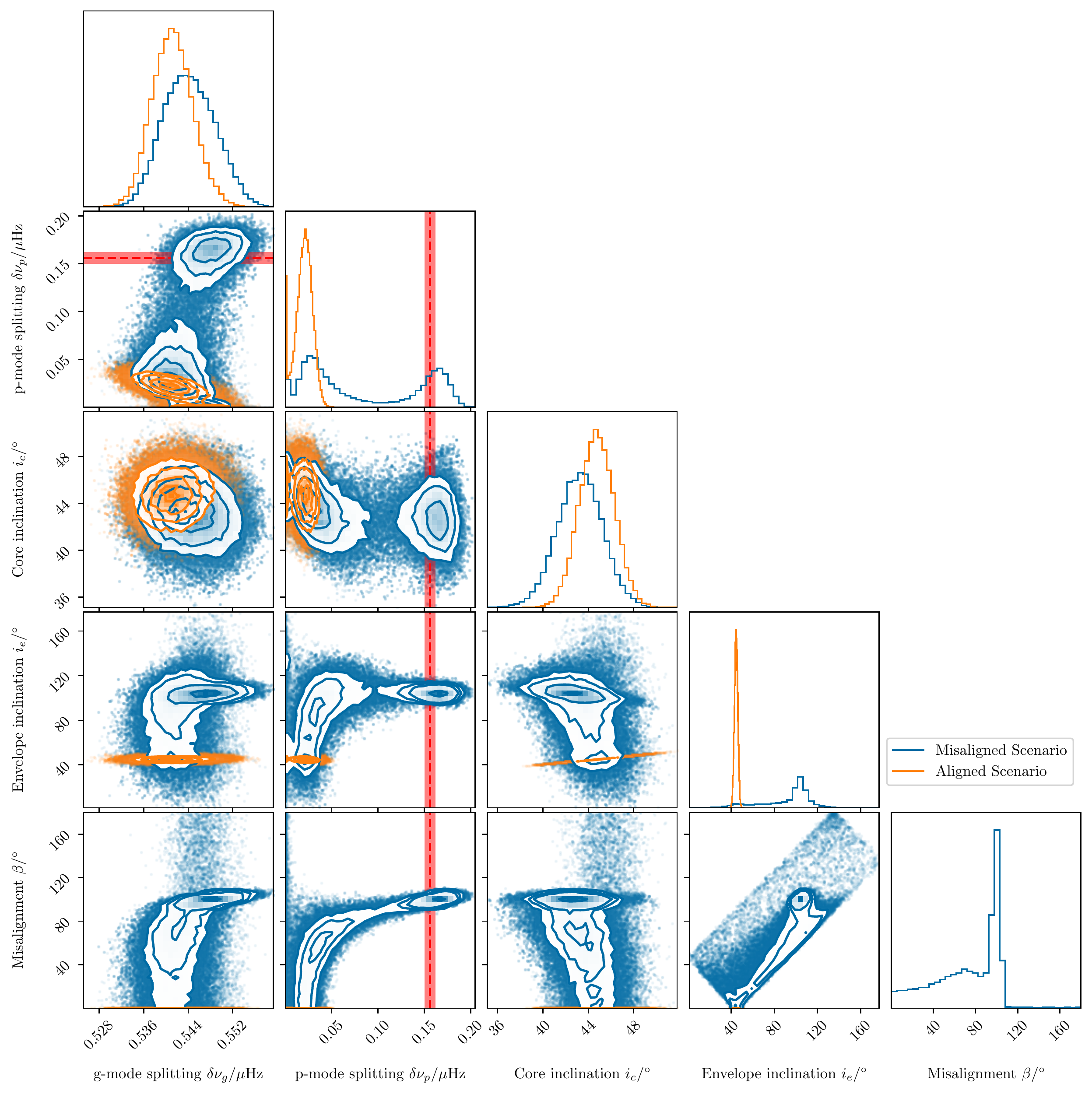}}{\node at (.648,.73){\includegraphics[width=.81\textwidth]{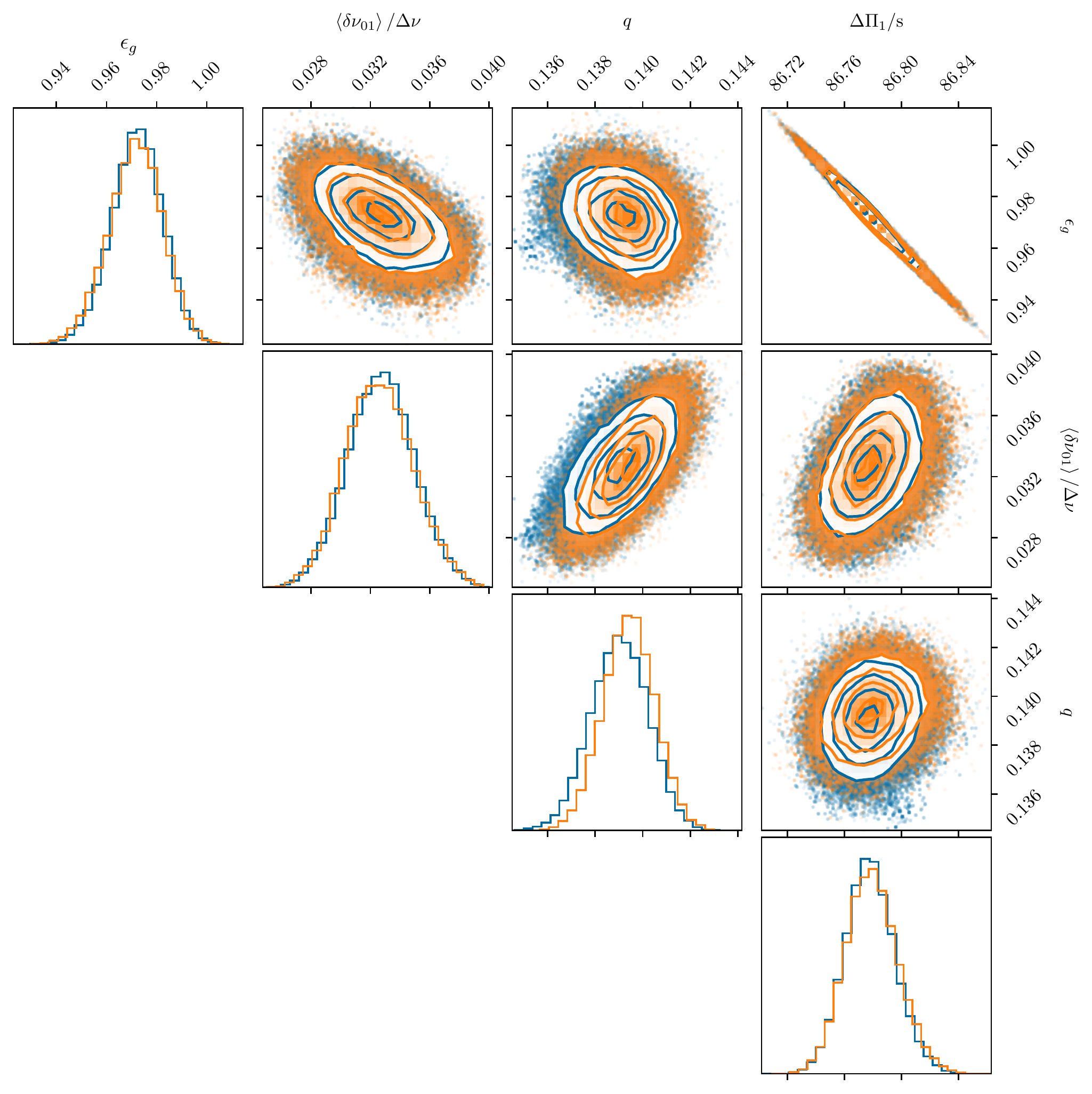}};}
    \caption{Posterior distribution for parameters of the PSD model. We compare a nominally misaligned scenario (blue) against a scenario where alignment between the core and envelope rotational axes is enforced (orange); in the latter, $\beta = 0$ by definition, and we therefore show no orange curves in the final row. The rotational parameters are shown in a corner plot originating in the bottom left, while the other mixed-mode parameters are shown in a separate one originating in the top right. The fitted envelope rotation rate in this aligned scenario can be seen to be systematically less than that returned from the misaligned case, as a result of the modifications to the avoided crossings that we examine in \autoref{sec:empirical}. The red dashed line, and shaded intervals, show the mean and uncertainties of the independently-measured photometric rotation rate of \cite{huber_stellar_2013}. This was not used as an input in our revised asteroseismic analysis, but confirms our selection of the rapidly-rotating mode in the misaligned posterior distribution.}
    \label{fig:posterior}
\end{figure*}

For more robust results (given the stochastic nature of mode excitation), we fit maximum-likelihood models against the entire power spectrum, and moreover infer Bayesian posterior distributions for the input parameters of this model to estimate uncertainties. We specify the prior distributions on these parameters in \cref{tab:prior}. Of the three angles shown in \cref{fig:triangle}, we sample \(\cos i_\text{core}\) uniformly (and thus isotropically) in the usual fashion, and place uniform priors on \(\varphi\) and \(\beta\), from which \(i_\text{core}\) and \(\lambda\) are computed using standard trigonometric identities. Note that to uninformatively sample the coordinates on the unit sphere defined by \(\varphi\) (azimuth) and \(\beta\) (colatitude), we ought in principle to be placing a uniform prior on \(\cos \beta\), so that the prior distribution is isotropic. However, we choose to sample \(\beta\) uniformly instead, reflecting a strong, and deliberately adversarial, prior assumption that the core and envelope of a star ought not to be misaligned. Likewise, we have deliberately imposed an adversarial, exponentially decaying prior distribution on both rotation rates, penalising rapid surface rotation in particular. We then draw samples from the posterior distribution using \texttt{dynesty} \citep{dynesty}, shown in blue in the corner plots of \cref{fig:posterior}. Draws from the posterior distribution are also shown with the faint gray curves in \cref{fig:ps}. For the purposes of model comparison, we repeat this procedure to fit models and draw samples in the aligned case (wherein we demand \(\beta = 0\), and set \(\varphi = 0\) also, since it is entirely unconstrained), shown with orange in \cref{fig:posterior}. We report the marginal posterior medians and \(\pm1\sigma\) quantiles of various model parameters for both the misaligned and aligned scenarios in \cref{tab:posterior}.

Prima facie, these posterior quantiles would suggest that expanding the parameter space to account for misalignment worsens the constraint on the envelope rotation rate, compared to the aligned scenario. However, inspecting the posterior distribution in \cref{fig:posterior} reveals that this is the result of multimodality. Specifically, there are two main peaks in the misaligned posterior distribution: one in which the core and envelope have the same inclination angle, and one where they are significantly misaligned. The former corresponds to a slow envelope rotation rate that is consistent with that fitted in the aligned scenario, while the latter corresponds to a much faster envelope rotation rate.

This fast-envelope-rotation mode in the posterior distribution contains both the maximum-likelihood solution, and also the majority of the posterior probability mass, despite being heavily disfavoured by our adversarial priors on the rotational splittings and on \(\beta\). Suggestively, its larger value of the envelope rotational splitting is also consistent with the photometric spot-modulation rotation rate of \(P_\text{rot} = 74\pm3\ \mathrm{d}\) derived in \citet{huber_stellar_2013}, shown with the red lines and shaded regions in \cref{fig:posterior}. We therefore argue that only this rapidly-envelope-rotating mode is representative of Kepler-56's true rotational orientation. Restricting our attention to it (truncating the posterior distribution at \(\delta\nu_p > 0.1\ \mu\mathrm{Hz}\)), the inclination angle of Kepler-56's envelope's rotational axis relative to the line of sight is \(i_\text{env} = (104 \pm 4)^\circ\), with a seismic rotation rate of \(\Omega_\text{env}/2\pi \sim \delta\nu_p = 0.16^{+0.01}_{-0.02}\ \mu\mathrm{Hz}\).

\begin{longtable}[]{@{}
  >{\raggedleft\arraybackslash}p{(\linewidth - 4\tabcolsep) * \real{0.3226}}
  >{\centering\arraybackslash}p{(\linewidth - 4\tabcolsep) * \real{0.3226}}
  >{\centering\arraybackslash}p{(\linewidth - 4\tabcolsep) * \real{0.3548}}@{}}
\caption{Inferred values of mixed-mode and rotational parameters of PSD model.\label{tab:posterior}}\tabularnewline
\toprule\noalign{}
\begin{minipage}[b]{\linewidth}\raggedleft
Parameter
\end{minipage} & \begin{minipage}[b]{\linewidth}\centering
Misaligned
\end{minipage} & \begin{minipage}[b]{\linewidth}\centering
Aligned
\end{minipage} \\
\midrule\noalign{}
\endfirsthead
\toprule\noalign{}
\begin{minipage}[b]{\linewidth}\raggedleft
Parameter
\end{minipage} & \begin{minipage}[b]{\linewidth}\centering
Misaligned
\end{minipage} & \begin{minipage}[b]{\linewidth}\centering
Aligned
\end{minipage} \\
\midrule\noalign{}
\endhead
\bottomrule\noalign{}
\endlastfoot
\(\Delta\Pi_1/\mathrm s\) & \(86.78 \pm 0.02\) & \(86.78 \pm 0.02\) \\
\(q\) & \(0.139 \pm 0.001\) & \(0.139 \pm 0.001\) \\
\(\epsilon_g\) & \(0.97 \pm 0.01\) & \(0.97 \pm 0.01\) \\
\(\left<\delta\nu_{01}/\Delta\nu\right>\) & \(0.033 \pm 0.002\) & \(0.033 \pm 0.002\) \\
\(\alpha_g\times 10^{8}\) & \(2.151\pm0.009\) & \(2.154^{+0.008}_{-0.009}\) \\
\midrule \(\delta\nu_g/\mu\mathrm{Hz}\) & \(0.544\pm 0.005\) & \(0.541\pm 0.004\) \\
\(\delta\nu_p/\mu\mathrm{Hz}\) & \(0.057^{+0.107}_{-0.035}\) & \(0.021^{+0.008}_{-0.010}\) \\
\(i_\text{core}/^\circ\) & \(43\pm 2\) & \(44.8 \pm 1.5\) \\
\(i_\text{env}/^\circ\) & \(101^{+8}_{-36}\) & ---- \\
\(\lambda/^\circ\) & \(60^{+30}_{-46}\) & ---- \\
\end{longtable}

We therefore also restrict our attention to only the maximum-likelihood estimate, which lies in the rapidly-rotating mode, for the purposes of model comparison against the aligned scenario. The misaligned model is significantly preferred, with a likelihood-ratio test statistic of \(\Lambda = 14.06\). This comes at the expense of introducing an additional two degrees of freedom compared to the nested aligned model where \(\lambda = 0\) and \(i_\text{env} = i_\text{core}\), so that the likelihood-ratio test yields a probability of \(p = 0.0009\) that this preference is by coincidence (i.e.~misalignment is preferred with \(> 3\sigma\) confidence). This being the case, and given the blinded consistency of the envelope rotation rate returned by this model with an independent measurement of the photometric rotation rate, we submit that the existing asteroseismic data for Kepler-56 \emph{already strongly suggest} that its core and envelope are heavily misaligned. Moreover, the inferred value for the envelope's inclination angle is such that the envelope is close to being seen equator-on. This is consistent with the envelope's rotational axis being aligned with the orbital axis of the inner transiting planets.

\subsection{Discussion: Spin-Orbit Interactions in Kepler-56}\label{sec:k56discussion}

Severe rotational misalignment between different parts of a star, of the kind inferred here, is not thought likely to spontaneously arise over the course of single-star evolution \citep[although e.g.][ propose a possible mechanism facilitated by internal gravity waves]{rogers_internal_2012, rogers_internal_2013}. Rather, tidal interactions have been suggested as a source of internal rotational misalignment in other contexts \citep[e.g.~in geophysics:][]{sikdar_differential_2023}. The presence of planets around Kepler-56 strongly suggests that this internal misalignment also originates from tidal interactions with its planets. We identify two main tidal timescales for physical effects relevant to determining the misalignment between the core and envelope:

\begin{enumerate}
\def\labelenumi{\arabic{enumi}.}
\tightlist
\item
  \emph{The realignment timescale between Kepler-56's convective envelope and its inner orbiting planets.} Kepler-56's rapid surface rotation may originate from tidal spin-up from its hot-Jupiter companions \citep[e.g.][]{brown_discrepancies_2014, maxted_comparison_2015, arevalo_further_2021}. These tidal interactions would also torque the envelope's rotational axis towards alignment with the planetary orbit normal \citep[e.g.][]{winn_hot_2010, albrecht_obliquities_2012, lai_tidal_2012, rogers_lin_2013, albrecht_preponderance_2021, patel_constraining_2022, rice_origins_2022}. For a sample of stars which have evolved off the main sequence, \citet{saunders_obliquity_2024} find that timescale of this tidal realignment may be at least 4 orders of magnitude shorter than that of orbital decay. They provide an upper limit of \(\sim 5\times 10^8\ \mathrm{yr}\) for the former timescale --- with the lower limit being essentially unconstrained --- but both of these tidal timescales scale with the stellar mean density. Kepler-56 is much less dense than the stars of their sample (by at least a factor of 8). Moreover, it is a red giant, so its realignment tidal quality factor is also likely much lower than the subgiants of \citet{saunders_obliquity_2024}, since much more of it is convectively unstable. Therefore, Kepler-56's realignment timescale should be at least an order of magnitude shorter than the subgiants of \citet{saunders_obliquity_2024}.
\item
  \emph{The apsidal/spin precession timescales around the total angular momentum vector of the system.} In addition to the two close-in planets reported in \citet{steffen_k56_2013}, Kepler-56 also hosts a more massive outer planet (\(M_\text{pl}\sin i_\text{pl} \sim 6 M_J\)) on a widely-separated, 2.7-year orbit \citep{otor_k56_2016, weiss_kgps_2024}, whose orbital angular momentum dominates the total angular momentum of the entire system. This causes both the stellar spin and planetary orbital angular momentum vectors to precess around this total angular momentum vector, as a result of gravitational torques exerted on tidal bulges. \citet{huber_stellar_2013} propose a scenario where the orbital precession occurs far more rapidly than spin precession in the absence of tidal dissipation, with the inner two planets precessing together. Repeating their numerical exercise using the \texttt{REBOUNDx} framework \citep[as formulated by][]{lu_selfconsistent_2023}, adopting their initial conditions and orbital properties, but updating \(a\), \(M_\mathrm{pl}\sin i_\mathrm{pl}\), and \(e\) for the outer planet per its radial-velocity orbital solution from \citet{weiss_kgps_2024}, yields estimates for its spin precession period of \(2\pi/\alpha \sim 1.45\times 10^6\ \mathrm{yr}\), and a nodal precession period of \(2\pi/g\sim 2\times 10^4\ \mathrm{yr}\).
\end{enumerate}

Other stellar-astrophysical processes may also torque the envelope and/or the core. These operate on different characteristic timescales:

\begin{enumerate}
\def\labelenumi{\arabic{enumi}.}
\setcounter{enumi}{2}
\tightlist
\item
  \emph{The internal realignment timescale between Kepler-56's convective envelope and its radiative core}. Radially differentially rotating cores and envelopes torque each other towards solid-body rotation \citep{aerts_angular_2019}; indeed, some angular momentum transport from the cores to the envelopes of red giants \citep[with transport timescales of \(\sim10^8\ \mathrm{yr}\), e.g.][]{fuller_slowing_2019} is known to be required to explain existing ensemble red-giant core and envelope rotation measurements \citep[e.g.][]{gehan_core_2018, li_asteroseismic_2024}. Since realignment between the core and envelope also entails them exerting torques on each other (albeit now no longer parallel to the direction of their angular momenta), we argue that the realignment timescale should also scale with the angular-momentum transport timescale, although potentially with further dependence on the misalignment angle. For a star of comparable mass to Kepler-56, this timescale is roughly comparable to the amount of time that is spent on the entire red giant branch \citep[e.g][]{eggenberger_constraining_2017, gehan_core_2018}.
\item
  \emph{The magnetic braking timescale of the envelope.} Our photometric rotation rate arises from surface features on Kepler-56, interpreted as being spots, producing brightness variations as they enter and leave the visible disk while its surface rotates. The presence of these spot features indicates magnetic activity, but these same physical processes also brake rotation. Applying the magnetic-braking prescription of \citeauthor{vansaders_fast_2013} \citetext{\citeyear{vansaders_fast_2013}; \citealp[and using the same calibrated parameters as in][ for red giants]{ong_zvrk_2024}} yields braking timescales of \(\gtrsim 3\times 10^7\ \mathrm{yr}\) at Kepler-56's current surface rotation rate.
\end{enumerate}

The relative durations of these timescales must be better understood for insight into how this misaligned configuration could have arisen, and how it might subsequently evolve. Lower bounds on the realignment timescale will be necessary to discriminate between different physical scenarios leading to internal misalignment. For example, if precession occurs much more rapidly than all dissipative (realignment and braking) timescales, we would anticipate the dynamics of the system to be dominated by gravitational torques. In the weakly nonadiabatic regime, the scenario of \citet{huber_stellar_2013} would eventually place the Kepler-56 system into Cassini State 2 \citep[e.g.][]{winn_obliquity_2005, fabrycky_obliquity_2007, millholland_formation_2020, su_dynamics_2022}, with the envelope spin, inner orbit, and total angular momentum vectors lying in the same plane. In this scenario, the gravitational torque acting on Kepler-56 would be exerted primarily on its envelope, while its core might either precess separately, or be treated as otherwise decoupled from this dynamical evolution.

Alternatively, if realignment had occurred rapidly upon evolution off the main sequence \citep[as suggested by][]{saunders_obliquity_2024}, we might expect the spin angular momentum of the envelope to have only recently been brought close to alignment with the inner planets. Otherwise, if we have caught Kepler-56's envelope in the act of tidal realignment, the envelope's rotation axis might be pointing at an intermediate direction to the core's axis and the orbital normal of the inner planets, in the process of realigning from one to the other. In this case, we would expect the angular momentum vectors of the core, envelope, and inner orbits to be coplanar, or close to coplanar. It may be possible to rule these scenarios out observationally by measuring the orbital obliquity \(\lambda_\text{pl}\) of the inner planets, through the Rossiter-McLaughlin effect. A coplanar configuration would require \(\lambda_\text{pl}\) to satisfy the constraint
\[
    \cos i_\text{core} \sin i_\text{env} \sin \lambda_\mathrm{pl} = \sin i_\text{core} \cos i_\text{env} \sin (\lambda+\lambda_\text{pl}),
\]
with the angles \(\lambda, i_\text{core}, i_\text{env}\) determined from asteroseismology. A planetary obliquity incompatible with this constraint would rule out such a recent, rapid realignment scenario. However, such measurements would be challenging for Kepler-56 in particular --- even planet c, with the greater transit depth of the two transiting planets, would only provide a signal of amplitude \(\sim2\ \mathrm{m\ s^{-1}}\) over a 12-hour-long transit.

Finally, if Kepler-56 had instead engulfed a planet formerly aligned with its present inner two \citep[e.g.][]{li_dynamics_2014}, this would also produce a rotational signature with the envelope brought closer into alignment with the planetary orbits than the core. Again, lower limits on both the tidal and internal realignment timescales will be required to place meaningful constraints on engulfment masses and radii under this scenario, given the observed rotation rates and misalignment angles.

\section{Discussion and Conclusion}\label{sec:discussion}

We have generalised the existing perturbative treatment of rotation in asteroseismology to accomodate a more general physical scenario in which the star is treated as a series of concentric mass shells, each of which is assigned a single rotational axis vector. We have then examined its consequences for a two-zone model of differential rotation, of the kind applied to red giant gravitoacoustic mixed modes, in which the core g-mode and envelope p-mode cavities are assumed to be only weakly coupled, both rotationally and pulsationally. In this regime, we obtain generalised expressions for the rotational splittings and implied multiplet inclination angles of gravitoacoustic mixed modes, both when near-degeneracy avoided crossings are neglected (\cref{eq:newgoupil}, which generalises the usual first-order expression \cref{eq:goupil}), and where they are fully accounted for (\cref{eq:tensor2,eq:tensorzeta,eq:tensorV}). These expressions indicate that in the presence of internal misalignment, the apparent inclination angle implied by the visiblity ratios between multiplet components will vary between multiplets, depending on their mixing fractions; this is our proposed observational signature of such misalignment.

Applying this revised formalism to a case study --- the misaligned planet host Kepler-56 --- indicates that its existing asteroseismic data set already strongly suggests its envelope and core rotate around different axes. The orientation of its envelope's rotational axis is moreover much closer to being seen equator-on, compared to the axis of its core. Thus, the former may well be aligned with the orbital angular momenta of its inner transiting planets, although confirming this will require the planetary obliquity angle also to be constrained.

Kepler-56 is one of only a very few planetary systems wherein significant spin-orbit misalignment has been previously reported in conjunction with multiple close-in, mutually aligned planets. Of these, only Kepler-56 and Kepler-129 \citep{zhang_long_2021} have had their rotational misalignment constrained using asteroseismology, and Kepler-56 is unique in being the only known such red-giant host star \citep{campante_spinorbit_2016}, permitting the use of gravitoacoustic mixed modes. Asteroseismology constrains averaged \emph{internal} rotation rates and orientations: this is not necessarily comparable to constraints on spin-orbit misalignment made using \emph{surface} observables like gravity darkening \citep[e.g.][]{ahlers_spinorbit_2015}, or the Rossiter-McLaughlin effect \citep{hjorth_backward_2021}. Over the last decade, the exoplanet community has used primarily surface stellar rotation to build up constraints on the spin-orbit angle over a large ensemble of extrasolar systems \citep[e.g.][]{winn_occurence_2015}. Our result suggests we may not yet rule out the possibility that a nontrivial fraction of the many known surface-spin-orbit-aligned planets are \emph{internally} spin-orbit-misaligned. Conversely, inspecting the internal rotation of known surface-spin-orbit-misaligned planets may also be fruitful for understanding the histories and physics of such misalignment. Asteroseismic rotational measurements remain our \emph{only} means of addressing these open astrophysical questions.

How might this reassessment of Kepler-56's present rotational orientation change our understanding of how its planets formed? Our preceding discussion supposes that Kepler-56's core and envelope were more rotationally coupled on the main sequence than presently --- and therefore that the present orientation of its core may preserve information about its rotation prior to any decoupling. If so, our separate constraints on the core and envelope do not immediately favour either dynamical \citep[e.g.][]{boue_compact_2014a, boue_compact_2014b, li_dynamics_2014} or disk-driven \citep[e.g.][]{spalding_early_2014, spalding_misalignment_2016, spalding_oblateness_2020} origins for any earlier misalignment, since our constraint on the core's orientation in particular is similar to that from \citet{huber_stellar_2013}. However, the fidelity with which the core's present orientation represents this earlier time is dependent on the timescale of internal realignment (not necessarily off the main sequence). As we have argued, this ought to scale with that of other, poorly-understood, angular-momentum transport processes also in operation in single stars. Moreover, we point out that the vast majority of Kepler-56's angular momentum is contained in the envelope, and so it too cannot be ignored when constraining these formation pathways. Thus, we further argue that incorporating these transport processes into models of the Kepler-56 system's rotational evolution will be necessary to discriminate between various hypotheses for how it formed, as their predictions must now produce consistency with our constraints on the orientations of both the core and the envelope. Conversely, this new relevance to planet formation \emph{increases} the urgency of better understanding these transport processes, which remains an elusive goal.

Follow-up spectroscopic radial-velocity measurements, through the Rossiter-McLaughlin effect, may better constrain these possible histories, as well as physical scenarios for how tidal realignment operates. Although our immediate prospects for doing this with Kepler-56 specifically are not encouraging, even our rough constraints on its internal spin configuration already suffice to provide some qualitative insight into the physics of any ongoing realignment. For example, realignment primarily modulated by gravity waves \citep[e.g.][]{rogers_internal_2012, zanazzi_damping_2024} would produce inside-out realignment --- rather than potential agreement between the envelope and orbital inclination angles suggested by our results --- since gravity waves are confined to the cores of red giants. This mechanism therefore likely does not dominate in Kepler-56. We leave more detailed examination of the physics behind spin-orbit realignment --- which ought to also be accompanied by more detailed stellar evolution and pulsation calculations fully accounting for the dependence of rotational splittings on internal structure --- to future work.

Aside from the applications of asteroseismology to studying star-planet interactions that our case study illustrates, we have also discussed modified expressions for the widths and distributions of power in rotational multiplets, in terms of internal rotation rates and inclination angles (\cref{eq:newgoupil}), which might change our interpretation of these observational features. In \autoref{sec:empirical}, we have shown that the specific manner in which these changes interact with existing techniques for asteroseismic rotational characterisation might yield systematic underestimates of envelope rotation rates. Depending on how widespread core-envelope misalignment is, this might imply that the true shape of the red-giant angular-momentum transport problem may differ, potentially significantly, from how our present observational characterisation represents it to be. In any case, stars exhibiting significant misalignment may not be desirable to include in the sample constraining this problem in the first place.

Revisiting existing envelope-rotation-rate measurements to accommodate misalignment, even if merely to diagnose its presence and exclude it from consideration, may therefore be important for uncontaminated observational constraints on angular-momentum transport. The primary logistical impediment to doing so is the fact that matrix construction that we extend in this work has so far not been widely applied for observational rotational characterisation of this kind. Instead, a JWKB asymptotic eigenvalue equation of the form
\[
    \tan \Theta_p \tan \Theta_g - q = 0 \label{eq:asy}
\]
is more commonly used \citep[e.g.][]{li_asteroseismic_2024, gehan_core_2018}, since it is both analytically simpler, and numerically more performant. A generalisation of the existing explicit one-to-one map between the rotating eigenvalue problem of \cref{eq:qhep1} and the asymptotic eigenvalue problem of \cref{eq:asy} \citep[e.g.][]{ong_rotation_2023}, to accommodate misalignment, may be essential to render computationally tractable such a population-level reassessment of red-giant differential rotation.

Finally, we submit that the deviations from azimuthal symmetry explored in this work underscore the increasing need for explicit 3D pulsation calculations. Semianalytic prescriptions of the kind provided here are only as good as the assumptions going into them. While the cores and envelopes of red giants are indeed well-approximated as both rotationally and pulsationally decoupled (\obb), this is not generally the case for main-sequence stars, and we might not necessarily expect the perturbative analysis here to hold well for them. Even perturbatively, one would expect the normal modes in those cases to at least be approximately of the form typically assumed in the Method of Lines \citep[e.g.][]{schiesser_numerical_2012}, where for the \(i^\mathrm{th}\) normal mode one assumes
\[
    f_i(r, \theta, \phi) = \sum_{\ell m} f_{i\ell m}(r) Y_\ell^m(\theta,\phi),
\]
rather than the usual variable-separable form associating each normal mode with only one spherical harmonic, so that the horizontal dependence of each normal mode is permitted to vary with radial position. Indeed, our use of the \(\pi/\gamma\)-mode decomposition here implicitly accommodates such variations of the horizontal dependence, albeit only in two separately-rotating zones. Full 3D pulsation calculations will be required to quantitatively validate theoretical treatments describing more general configurations.

\section*{Acknowledgements}\label{acknowledgements}
\addcontentsline{toc}{section}{Acknowledgements}

I am grateful for the anonymous referee's constructive criticism; it significantly improved the presentation of this work. I thank Jennifer van Saders, for productive discussions that ultimately led to this work; Daniel Huber, for insightful remarks, and for kindly providing access to the full posterior distributions from his original analysis of Kepler-56; Sarbani Basu, Margarida Cunha, and Nicholas Rui, for constructive feedback on preliminary versions of this work; and Amaury Triaud, Malena Rice, Fei Dai, and Josh Winn, for thoughtful comments from the exoplanet community. I also thank Joanne Tan for meaningful assistance in getting my analysis code working. I acknowledge support from NASA through the NASA Hubble Fellowship grant HST-HF2-51517.001, awarded by STScI. STScI is operated by the Association of Universities for Research in Astronomy, Incorporated, under NASA contract NAS5-26555.

\facilities{\emph{Kepler}}

\software{NumPy \citep{numpy}, SciPy stack \citep{scipy}, AstroPy \citep[astropy:2018]{astropy:2013}, Pandas \citep{pandas}, \mesa~\citep{mesa_paper_1, mesa_paper_2, mesa_paper_3, mesa_paper_4, mesa_paper_5}, \gyre~\citep{townsend_gyre_2013}, \texttt{jax} \citep{jax}, \texttt{dynesty} \citep{dynesty}, \texttt{PBJam}~\citep{pbjam}, \texttt{reggae}~\citep{ong_reggae_2024}, \texttt{REBOUND}~\citep{rebound}, \texttt{REBOUNDx}~\citep{reboundx}.}

The analysis code and \mesa~models used in this work are made available on Zenodo at \dataset[doi:10.5281/zenodo.14512016]{\doi{10.5281/zenodo.14512016}}.

\bibliography{biblio.bib,custom.bib}

\end{document}

%% file: macros.tex
\usepackage{CJKutf8}
\usepackage{newunicodechar}

\newcommand{\numax}{\ensuremath{{\nu_\text{max}}}}

\newcommand{\ii}{{\ensuremath{\mathrm{i}}}}

\newcommand\mesa{{\textsc{mesa}}}
\newcommand\gyre{{\textsc{gyre}}}
\newcommand\Kepler{{\textit{Kepler}}}

\defcitealias{ong_semianalytic_2020}{OB20}
\newcommand\ob{{\citetalias{ong_semianalytic_2020}}}
\defcitealias{ong_rotation_2022}{OBB22}
\newcommand\obb{{\citetalias{ong_rotation_2022}}}
\defcitealias{ong_rotation_2023}{OG23}

\newcommand{\chinesename}{{\begin{CJK}{UTF8}{gbsn}(王加冕)\end{CJK}}}

\DeclareRobustCommand{\okina}{%
  \raisebox{\dimexpr\fontcharht\font`A-\height}{%
    \scalebox{0.8}{`}%
  }%
}
\newunicodechar{ʻ}{\okina}

\graphicspath{{./},{./figures/}}


%% file: oblique.bbl
\begin{thebibliography}{}
\expandafter\ifx\csname natexlab\endcsname\relax\def\natexlab#1{#1}\fi
\providecommand{\url}[1]{\href{#1}{#1}}
\providecommand{\mhref}[2]{\href{#1}{\color{magenta}#2}}
\providecommand{\dodoi}[1]{doi:~\href{http://doi.org/#1}{\nolinkurl{#1}}}
\providecommand{\doeprint}[1]{\href{http://ascl.net/#1}{\nolinkurl{http://ascl.net/#1}}}
\providecommand{\doarXiv}[1]{\href{https://arxiv.org/abs/#1}{\nolinkurl{https://arxiv.org/abs/#1}}}

\bibitem[{{Aerts} {et~al.}(2010){Aerts}, {Christensen-Dalsgaard}, \&
  {Kurtz}}]{aertsbook}
{Aerts}, C., {Christensen-Dalsgaard}, J., \& {Kurtz}, D.~W. 2010,
  {Asteroseismology} (Berlin: Springer)

\bibitem[{{Aerts} {et~al.}(2019){Aerts}, {Mathis}, \&
  {Rogers}}]{aerts_angular_2019}
{Aerts}, C., {Mathis}, S., \& {Rogers}, T.~M. 2019,
  {\mhref{http://doi.org/10.1146/annurev-astro-091918-104359}{\araa}},
  {\href{https://ui.adsabs.harvard.edu/abs/2019ARA&A..57...35A}{57}}{\href{https://ui.adsabs.harvard.edu/abs/2019ARA&A..57...35A}{,
  35}}

\bibitem[{{Ahlborn} {et~al.}(2022){Ahlborn}, {Bellinger}, {Hekker}, {Basu}, \&
  {Mokrytska}}]{ahlborn_improved_2022}
{Ahlborn}, F., {Bellinger}, E.~P., {Hekker}, S., {Basu}, S., \& {Mokrytska}, D.
  2022, {\mhref{http://doi.org/10.1051/0004-6361/202142510}{\aap}},
  {\href{https://ui.adsabs.harvard.edu/abs/2022A&A...668A..98A}{668}}{\href{https://ui.adsabs.harvard.edu/abs/2022A&A...668A..98A}{,
  A98}}

\bibitem[{{Ahlers} {et~al.}(2015){Ahlers}, {Barnes}, \&
  {Barnes}}]{ahlers_spinorbit_2015}
{Ahlers}, J.~P., {Barnes}, J.~W., \& {Barnes}, R. 2015,
  {\mhref{http://doi.org/10.1088/0004-637X/814/1/67}{\apj}},
  {\href{https://ui.adsabs.harvard.edu/abs/2015ApJ...814...67A}{814}}{\href{https://ui.adsabs.harvard.edu/abs/2015ApJ...814...67A}{,
  67}}

\bibitem[{{Albrecht} {et~al.}(2012){Albrecht}, {Winn}, {Johnson}, {Howard},
  {Marcy}, {Butler}, {Arriagada}, {Crane}, {Shectman}, {Thompson}, {Hirano},
  {Bakos}, \& {Hartman}}]{albrecht_obliquities_2012}
{Albrecht}, S., {Winn}, J.~N., {Johnson}, J.~A., {et~al.} 2012,
  {\mhref{http://doi.org/10.1088/0004-637X/757/1/18}{\apj}},
  {\href{https://ui.adsabs.harvard.edu/abs/2012ApJ...757...18A}{757}}{\href{https://ui.adsabs.harvard.edu/abs/2012ApJ...757...18A}{,
  18}}

\bibitem[{{Albrecht} {et~al.}(2021){Albrecht}, {Marcussen}, {Winn}, {Dawson},
  \& {Knudstrup}}]{albrecht_preponderance_2021}
{Albrecht}, S.~H., {Marcussen}, M.~L., {Winn}, J.~N., {Dawson}, R.~I., \&
  {Knudstrup}, E. 2021,
  {\mhref{http://doi.org/10.3847/2041-8213/ac0f03}{\apjl}},
  {\href{https://ui.adsabs.harvard.edu/abs/2021ApJ...916L...1A}{916}}{\href{https://ui.adsabs.harvard.edu/abs/2021ApJ...916L...1A}{,
  L1}}

\bibitem[{{Astropy Collaboration} {et~al.}(2013){Astropy Collaboration},
  {Robitaille}, {Tollerud}, {Greenfield}, {Droettboom}, {Bray}, {Aldcroft},
  {Davis}, {Ginsburg}, {Price-Whelan}, {Kerzendorf}, {Conley}, {Crighton},
  {Barbary}, {Muna}, {Ferguson}, {Grollier}, {Parikh}, {Nair}, {Unther},
  {Deil}, {Woillez}, {Conseil}, {Kramer}, {Turner}, {Singer}, {Fox}, {Weaver},
  {Zabalza}, {Edwards}, {Azalee Bostroem}, {Burke}, {Casey}, {Crawford},
  {Dencheva}, {Ely}, {Jenness}, {Labrie}, {Lim}, {Pierfederici}, {Pontzen},
  {Ptak}, {Refsdal}, {Servillat}, \& {Streicher}}]{astropy:2013}
{Astropy Collaboration}, {Robitaille}, T.~P., {Tollerud}, E.~J., {et~al.} 2013,
  {\mhref{http://doi.org/10.1051/0004-6361/201322068}{\aap}},
  {\href{https://ui.adsabs.harvard.edu/abs/2013A&A...558A..33A}{558}}{\href{https://ui.adsabs.harvard.edu/abs/2013A&A...558A..33A}{,
  A33}}

\bibitem[{{Beck} {et~al.}(2012){Beck}, {Montalban}, {Kallinger}, {De Ridder},
  {Aerts}, {Garc{\'\i}a}, {Hekker}, {Dupret}, {Mosser}, {Eggenberger},
  {Stello}, {Elsworth}, {Frandsen}, {Carrier}, {Hillen}, {Gruberbauer},
  {Christensen-Dalsgaard}, {Miglio}, {Valentini}, {Bedding}, {Kjeldsen},
  {Girouard}, {Hall}, \& {Ibrahim}}]{beck_fast_2012}
{Beck}, P.~G., {Montalban}, J., {Kallinger}, T., {et~al.} 2012,
  {\mhref{http://doi.org/10.1038/nature10612}{\nat}},
  {\href{https://ui.adsabs.harvard.edu/abs/2012Natur.481...55B}{481}}{\href{https://ui.adsabs.harvard.edu/abs/2012Natur.481...55B}{,
  55}}

\bibitem[{{Belkacem} {et~al.}(2015){Belkacem}, {Marques}, {Goupil}, {Mosser},
  {Sonoi}, {Ouazzani}, {Dupret}, {Mathis}, \&
  {Grosjean}}]{belkacem_angular_2015}
{Belkacem}, K., {Marques}, J.~P., {Goupil}, M.~J., {et~al.} 2015,
  {\mhref{http://doi.org/10.1051/0004-6361/201526043}{\aap}},
  {\href{https://ui.adsabs.harvard.edu/abs/2015A&A...579A..31B}{579}}{\href{https://ui.adsabs.harvard.edu/abs/2015A&A...579A..31B}{,
  A31}}

\bibitem[{{Benomar} {et~al.}(2014){Benomar}, {Belkacem}, {Bedding}, {Stello},
  {Di Mauro}, {Ventura}, {Mosser}, {Goupil}, {Samadi}, \&
  {Garcia}}]{benomar_evolved_2014}
{Benomar}, O., {Belkacem}, K., {Bedding}, T.~R., {et~al.} 2014,
  {\mhref{http://doi.org/10.1088/2041-8205/781/2/L29}{\apjl}},
  {\href{https://ui.adsabs.harvard.edu/abs/2014ApJ...781L..29B}{781}}{\href{https://ui.adsabs.harvard.edu/abs/2014ApJ...781L..29B}{,
  L29}}

\bibitem[{{Bou{\'e}} \& {Fabrycky}(2014{\natexlab{a}})}]{boue_compact_2014a}
{Bou{\'e}}, G., \& {Fabrycky}, D.~C. 2014{\natexlab{a}},
  {\mhref{http://doi.org/10.1088/0004-637X/789/2/110}{\apj}},
  {\href{https://ui.adsabs.harvard.edu/abs/2014ApJ...789..110B}{789}}{\href{https://ui.adsabs.harvard.edu/abs/2014ApJ...789..110B}{,
  110}}

\bibitem[{{Bou{\'e}} \& {Fabrycky}(2014{\natexlab{b}})}]{boue_compact_2014b}
---. 2014{\natexlab{b}},
  {\mhref{http://doi.org/10.1088/0004-637X/789/2/111}{\apj}},
  {\href{https://ui.adsabs.harvard.edu/abs/2014ApJ...789..111B}{789}}{\href{https://ui.adsabs.harvard.edu/abs/2014ApJ...789..111B}{,
  111}}

\bibitem[{Bradbury {et~al.}(2018)Bradbury, Frostig, Hawkins, Johnson, Leary,
  Maclaurin, Necula, Paszke, Vander{P}las, Wanderman-{M}ilne, \& Zhang}]{jax}
Bradbury, J., Frostig, R., Hawkins, P., {et~al.} 2018, {JAX}: composable
  transformations of {P}ython+{N}um{P}y programs, 0.3.13.
\newblock \url{http://github.com/google/jax}

\bibitem[{{Brown}(2014)}]{brown_discrepancies_2014}
{Brown}, D.~J.~A. 2014, {\mhref{http://doi.org/10.1093/mnras/stu950}{\mnras}},
  {\href{https://ui.adsabs.harvard.edu/abs/2014MNRAS.442.1844B}{442}}{\href{https://ui.adsabs.harvard.edu/abs/2014MNRAS.442.1844B}{,
  1844}}

\bibitem[{{Campante} {et~al.}(2016){Campante}, {Lund}, {Kuszlewicz}, {Davies},
  {Chaplin}, {Albrecht}, {Winn}, {Bedding}, {Benomar}, {Bossini}, {Handberg},
  {Santos}, {Van Eylen}, {Basu}, {Christensen-Dalsgaard}, {Elsworth}, {Hekker},
  {Hirano}, {Huber}, {Karoff}, {Kjeldsen}, {Lundkvist}, {North}, {Silva
  Aguirre}, {Stello}, \& {White}}]{campante_spinorbit_2016}
{Campante}, T.~L., {Lund}, M.~N., {Kuszlewicz}, J.~S., {et~al.} 2016,
  {\mhref{http://doi.org/10.3847/0004-637X/819/1/85}{\apj}},
  {\href{https://ui.adsabs.harvard.edu/abs/2016ApJ...819...85C}{819}}{\href{https://ui.adsabs.harvard.edu/abs/2016ApJ...819...85C}{,
  85}}

\bibitem[{{Ceillier} {et~al.}(2017){Ceillier}, {Tayar}, {Mathur}, {Salabert},
  {Garc{\'\i}a}, {Stello}, {Pinsonneault}, {van Saders}, {Beck}, \&
  {Bloemen}}]{ceillier_surface_2017}
{Ceillier}, T., {Tayar}, J., {Mathur}, S., {et~al.} 2017,
  {\mhref{http://doi.org/10.1051/0004-6361/201629884}{\aap}},
  {\href{https://ui.adsabs.harvard.edu/abs/2017A&A...605A.111C}{605}}{\href{https://ui.adsabs.harvard.edu/abs/2017A&A...605A.111C}{,
  A111}}

\bibitem[{{De} {et~al.}(2023){De}, {MacLeod}, {Karambelkar}, {Jencson},
  {Chakrabarty}, {Conroy}, {Dekany}, {Eilers}, {Graham}, {Hillenbrand}, {Kara},
  {Kasliwal}, {Kulkarni}, {Lau}, {Loeb}, {Masci}, {Medford}, {Meisner},
  {Patel}, {Quiroga-Nu{\~n}ez}, {Riddle}, {Rusholme}, {Simcoe}, {Sjouwerman},
  {Teague}, \& {Vanderburg}}]{de_infrared_2023}
{De}, K., {MacLeod}, M., {Karambelkar}, V., {et~al.} 2023,
  {\mhref{http://doi.org/10.1038/s41586-023-05842-x}{\nat}},
  {\href{https://ui.adsabs.harvard.edu/abs/2023Natur.617...55D}{617}}{\href{https://ui.adsabs.harvard.edu/abs/2023Natur.617...55D}{,
  55}}

\bibitem[{{Deheuvels} {et~al.}(2023){Deheuvels}, {Li}, {Ballot}, \&
  {Ligni{\`e}res}}]{deheuvels_strong_2023}
{Deheuvels}, S., {Li}, G., {Ballot}, J., \& {Ligni{\`e}res}, F. 2023,
  {\mhref{http://doi.org/10.1051/0004-6361/202245282}{\aap}},
  {\href{https://ui.adsabs.harvard.edu/abs/2023A&A...670L..16D}{670}}{\href{https://ui.adsabs.harvard.edu/abs/2023A&A...670L..16D}{,
  L16}}

\bibitem[{{Deheuvels} {et~al.}(2017){Deheuvels}, {Ouazzani}, \&
  {Basu}}]{deheuvels_near_2017}
{Deheuvels}, S., {Ouazzani}, R.~M., \& {Basu}, S. 2017,
  {\mhref{http://doi.org/10.1051/0004-6361/201730786}{\aap}},
  {\href{https://ui.adsabs.harvard.edu/abs/2017A&A...605A..75D}{605}}{\href{https://ui.adsabs.harvard.edu/abs/2017A&A...605A..75D}{,
  A75}}

\bibitem[{{Deheuvels} {et~al.}(2012){Deheuvels}, {Garc{\'\i}a}, {Chaplin},
  {Basu}, {Antia}, {Appourchaux}, {Benomar}, {Davies}, {Elsworth}, {Gizon},
  {Goupil}, {Reese}, {Regulo}, {Schou}, {Stahn}, {Casagrande},
  {Christensen-Dalsgaard}, {Fischer}, {Hekker}, {Kjeldsen}, {Mathur}, {Mosser},
  {Pinsonneault}, {Valenti}, {Christiansen}, {Kinemuchi}, \&
  {Mullally}}]{deheuvels_seismic_2012}
{Deheuvels}, S., {Garc{\'\i}a}, R.~A., {Chaplin}, W.~J., {et~al.} 2012,
  {\mhref{http://doi.org/10.1088/0004-637X/756/1/19}{\apj}},
  {\href{https://ui.adsabs.harvard.edu/abs/2012ApJ...756...19D}{756}}{\href{https://ui.adsabs.harvard.edu/abs/2012ApJ...756...19D}{,
  19}}

\bibitem[{{Deheuvels} {et~al.}(2014){Deheuvels}, {Do{\u{g}}an}, {Goupil},
  {Appourchaux}, {Benomar}, {Bruntt}, {Campante}, {Casagrande}, {Ceillier},
  {Davies}, {De Cat}, {Fu}, {Garc{\'\i}a}, {Lobel}, {Mosser}, {Reese},
  {Regulo}, {Schou}, {Stahn}, {Thygesen}, {Yang}, {Chaplin},
  {Christensen-Dalsgaard}, {Eggenberger}, {Gizon}, {Mathis},
  {Molenda-{\.Z}akowicz}, \& {Pinsonneault}}]{deheuvels_seismic_2014}
{Deheuvels}, S., {Do{\u{g}}an}, G., {Goupil}, M.~J., {et~al.} 2014,
  {\mhref{http://doi.org/10.1051/0004-6361/201322779}{\aap}},
  {\href{https://ui.adsabs.harvard.edu/abs/2014A&A...564A..27D}{564}}{\href{https://ui.adsabs.harvard.edu/abs/2014A&A...564A..27D}{,
  A27}}

\bibitem[{{Dr{\'e}au} {et~al.}(2020){Dr{\'e}au}, {Cunha}, {Vrard}, \&
  {Avelino}}]{dreau_dipolar_2020}
{Dr{\'e}au}, G., {Cunha}, M.~S., {Vrard}, M., \& {Avelino}, P.~P. 2020,
  {\mhref{http://doi.org/10.1093/mnras/staa1981}{\mnras}},
  {\href{https://ui.adsabs.harvard.edu/abs/2020MNRAS.497.1008D}{497}}{\href{https://ui.adsabs.harvard.edu/abs/2020MNRAS.497.1008D}{,
  1008}}

\bibitem[{{Dupret} {et~al.}(2009){Dupret}, {Belkacem}, {Samadi}, {Montalban},
  {Moreira}, {Miglio}, {Godart}, {Ventura}, {Ludwig}, {Grigahc{\`e}ne},
  {Goupil}, {Noels}, \& {Caffau}}]{dupret_theoretical_2009}
{Dupret}, M.~A., {Belkacem}, K., {Samadi}, R., {et~al.} 2009,
  {\mhref{http://doi.org/10.1051/0004-6361/200911713}{\aap}},
  {\href{https://ui.adsabs.harvard.edu/abs/2009A&A...506...57D}{506}}{\href{https://ui.adsabs.harvard.edu/abs/2009A&A...506...57D}{,
  57}}

\bibitem[{{Eggenberger} {et~al.}(2017){Eggenberger}, {Lagarde}, {Miglio},
  {Montalb{\'a}n}, {Ekstr{\"o}m}, {Georgy}, {Meynet}, {Salmon}, {Ceillier},
  {Garc{\'\i}a}, {Mathis}, {Deheuvels}, {Maeder}, {den Hartogh}, \&
  {Hirschi}}]{eggenberger_constraining_2017}
{Eggenberger}, P., {Lagarde}, N., {Miglio}, A., {et~al.} 2017,
  {\mhref{http://doi.org/10.1051/0004-6361/201629459}{\aap}},
  {\href{https://ui.adsabs.harvard.edu/abs/2017A&A...599A..18E}{599}}{\href{https://ui.adsabs.harvard.edu/abs/2017A&A...599A..18E}{,
  A18}}

\bibitem[{{Fabrycky} {et~al.}(2007){Fabrycky}, {Johnson}, \&
  {Goodman}}]{fabrycky_obliquity_2007}
{Fabrycky}, D.~C., {Johnson}, E.~T., \& {Goodman}, J. 2007,
  {\mhref{http://doi.org/10.1086/519075}{\apj}},
  {\href{https://ui.adsabs.harvard.edu/abs/2007ApJ...665..754F}{665}}{\href{https://ui.adsabs.harvard.edu/abs/2007ApJ...665..754F}{,
  754}}

\bibitem[{{Fuller} {et~al.}(2019){Fuller}, {Piro}, \&
  {Jermyn}}]{fuller_slowing_2019}
{Fuller}, J., {Piro}, A.~L., \& {Jermyn}, A.~S. 2019,
  {\mhref{http://doi.org/10.1093/mnras/stz514}{\mnras}},
  {\href{https://ui.adsabs.harvard.edu/abs/2019MNRAS.485.3661F}{485}}{\href{https://ui.adsabs.harvard.edu/abs/2019MNRAS.485.3661F}{,
  3661}}

\bibitem[{{Garc{\'\i}a} {et~al.}(2011){Garc{\'\i}a}, {Hekker}, {Stello},
  {Guti{\'e}rrez-Soto}, {Handberg}, {Huber}, {Karoff}, {Uytterhoeven},
  {Appourchaux}, {Chaplin}, {Elsworth}, {Mathur}, {Ballot},
  {Christensen-Dalsgaard}, {Gilliland}, {Houdek}, {Jenkins}, {Kjeldsen},
  {McCauliff}, {Metcalfe}, {Middour}, {Molenda-Zakowicz}, {Monteiro}, {Smith},
  \& {Thompson}}]{kepseismic1}
{Garc{\'\i}a}, R.~A., {Hekker}, S., {Stello}, D., {et~al.} 2011,
  {\mhref{http://doi.org/10.1111/j.1745-3933.2011.01042.x}{\mnras}},
  {\href{https://ui.adsabs.harvard.edu/abs/2011MNRAS.414L...6G}{414}}{\href{https://ui.adsabs.harvard.edu/abs/2011MNRAS.414L...6G}{,
  L6}}

\bibitem[{{Gaulme} {et~al.}(2020){Gaulme}, {Jackiewicz}, {Spada}, {Chojnowski},
  {Mosser}, {McKeever}, {Hedlund}, {Vrard}, {Benbakoura}, \&
  {Damiani}}]{gaulme_active_2020}
{Gaulme}, P., {Jackiewicz}, J., {Spada}, F., {et~al.} 2020,
  {\mhref{http://doi.org/10.1051/0004-6361/202037781}{\aap}},
  {\href{https://ui.adsabs.harvard.edu/abs/2020A&A...639A..63G}{639}}{\href{https://ui.adsabs.harvard.edu/abs/2020A&A...639A..63G}{,
  A63}}

\bibitem[{{Gehan} {et~al.}(2021){Gehan}, {Mosser}, {Michel}, \&
  {Cunha}}]{gehan_automated_2021}
{Gehan}, C., {Mosser}, B., {Michel}, E., \& {Cunha}, M.~S. 2021,
  {\mhref{http://doi.org/10.1051/0004-6361/202039285}{\aap}},
  {\href{https://ui.adsabs.harvard.edu/abs/2021A&A...645A.124G}{645}}{\href{https://ui.adsabs.harvard.edu/abs/2021A&A...645A.124G}{,
  A124}}

\bibitem[{{Gehan} {et~al.}(2018){Gehan}, {Mosser}, {Michel}, {Samadi}, \&
  {Kallinger}}]{gehan_core_2018}
{Gehan}, C., {Mosser}, B., {Michel}, E., {Samadi}, R., \& {Kallinger}, T. 2018,
  {\mhref{http://doi.org/10.1051/0004-6361/201832822}{\aap}},
  {\href{https://ui.adsabs.harvard.edu/abs/2018A&A...616A..24G}{616}}{\href{https://ui.adsabs.harvard.edu/abs/2018A&A...616A..24G}{,
  A24}}

\bibitem[{{Gizon} \& {Solanki}(2003)}]{gizon_inclination_2003}
{Gizon}, L., \& {Solanki}, S.~K. 2003,
  {\mhref{http://doi.org/10.1086/374715}{\apj}},
  {\href{https://ui.adsabs.harvard.edu/abs/2003ApJ...589.1009G}{589}}{\href{https://ui.adsabs.harvard.edu/abs/2003ApJ...589.1009G}{,
  1009}}

\bibitem[{{Gough}(1993)}]{gough_linear_1993}
{Gough}, D.~O. 1993, in Astrophysical Fluid Dynamics - Les Houches 1987, ed.
  J.-P. {Zahn} \& J.~{Zinn-Justin} (Amsterdam: North-Holland),
  {\href{http://adsabs.harvard.edu/abs/1993afd..conf..399G}{399--560}}

\bibitem[{{Gough}(2007)}]{gough_jwkb_2007}
{Gough}, D.~O. 2007,
  {\mhref{http://doi.org/10.1002/asna.200610730}{Astronomische Nachrichten}},
  {\href{https://ui.adsabs.harvard.edu/abs/2007AN....328..273G}{328}}{\href{https://ui.adsabs.harvard.edu/abs/2007AN....328..273G}{,
  273}}

\bibitem[{{Gough} \& {Kosovichev}(1993)}]{gough_possible_1993}
{Gough}, D.~O., \& {Kosovichev}, A.~G. 1993, in Astronomical Society of the
  Pacific Conference Series, Vol.~40, IAU Colloq. 137: Inside the Stars, ed.
  W.~W. {Weiss} \& A.~{Baglin},
  {\href{https://ui.adsabs.harvard.edu/abs/1993ASPC...40..566G}{566}}

\bibitem[{{Gough} \& {Thompson}(1990)}]{gough_rotation_1990}
{Gough}, D.~O., \& {Thompson}, M.~J. 1990,
  {\mhref{http://doi.org/10.1093/mnras/242.1.25}{\mnras}},
  {\href{https://ui.adsabs.harvard.edu/abs/1990MNRAS.242...25G}{242}}{\href{https://ui.adsabs.harvard.edu/abs/1990MNRAS.242...25G}{,
  25}}

\bibitem[{{Goupil} {et~al.}(2013){Goupil}, {Mosser}, {Marques}, {Ouazzani},
  {Belkacem}, {Lebreton}, \& {Samadi}}]{goupil_seismic_2013}
{Goupil}, M.~J., {Mosser}, B., {Marques}, J.~P., {et~al.} 2013,
  {\mhref{http://doi.org/10.1051/0004-6361/201220266}{\aap}},
  {\href{https://ui.adsabs.harvard.edu/abs/2013A&A...549A..75G}{549}}{\href{https://ui.adsabs.harvard.edu/abs/2013A&A...549A..75G}{,
  A75}}

\bibitem[{{Hall} {et~al.}(2021){Hall}, {Davies}, {van Saders}, {Nielsen},
  {Lund}, {Chaplin}, {Garc{\'\i}a}, {Amard}, {Breimann}, {Khan}, {See}, \&
  {Tayar}}]{hall_weakened_2021}
{Hall}, O.~J., {Davies}, G.~R., {van Saders}, J., {et~al.} 2021,
  {\mhref{http://doi.org/10.1038/s41550-021-01335-x}{Nature Astronomy}},
  {\href{https://ui.adsabs.harvard.edu/abs/2021NatAs...5..707H}{5}}{\href{https://ui.adsabs.harvard.edu/abs/2021NatAs...5..707H}{,
  707}}

\bibitem[{{Harris} {et~al.}(2020){Harris}, {Millman}, {van der Walt},
  {Gommers}, {Virtanen}, {Cournapeau}, {Wieser}, {Taylor}, {Berg}, {Smith},
  {Kern}, {Picus}, {Hoyer}, {van Kerkwijk}, {Brett}, {Haldane}, {del R{\'\i}o},
  {Wiebe}, {Peterson}, {G{\'e}rard-Marchant}, {Sheppard}, {Reddy}, {Weckesser},
  {Abbasi}, {Gohlke}, \& {Oliphant}}]{numpy}
{Harris}, C.~R., {Millman}, K.~J., {van der Walt}, S.~J., {et~al.} 2020,
  {\mhref{http://doi.org/10.1038/s41586-020-2649-2}{\nat}},
  {\href{https://ui.adsabs.harvard.edu/abs/2020Natur.585..357H}{585}}{\href{https://ui.adsabs.harvard.edu/abs/2020Natur.585..357H}{,
  357}}

\bibitem[{{Hjorth} {et~al.}(2021){Hjorth}, {Albrecht}, {Hirano}, {Winn},
  {Dawson}, {Zanazzi}, {Knudstrup}, \& {Sato}}]{hjorth_backward_2021}
{Hjorth}, M., {Albrecht}, S., {Hirano}, T., {et~al.} 2021,
  {\mhref{http://doi.org/10.1073/pnas.2017418118}{Proceedings of the National
  Academy of Science}},
  {\href{https://ui.adsabs.harvard.edu/abs/2021PNAS..11817418H}{118}}{\href{https://ui.adsabs.harvard.edu/abs/2021PNAS..11817418H}{,
  e2017418118}}

\bibitem[{{Huber} {et~al.}(2013){Huber}, {Carter}, {Barbieri}, {Miglio},
  {Deck}, {Fabrycky}, {Montet}, {Buchhave}, {Chaplin}, {Hekker},
  {Montalb{\'a}n}, {Sanchis-Ojeda}, {Basu}, {Bedding}, {Campante},
  {Christensen-Dalsgaard}, {Elsworth}, {Stello}, {Arentoft}, {Ford},
  {Gilliland}, {Handberg}, {Howard}, {Isaacson}, {Johnson}, {Karoff},
  {Kawaler}, {Kjeldsen}, {Latham}, {Lund}, {Lundkvist}, {Marcy}, {Metcalfe},
  {Silva Aguirre}, \& {Winn}}]{huber_stellar_2013}
{Huber}, D., {Carter}, J.~A., {Barbieri}, M., {et~al.} 2013,
  {\mhref{http://doi.org/10.1126/science.1242066}{Science}},
  {\href{https://ui.adsabs.harvard.edu/abs/2013Sci...342..331H}{342}}{\href{https://ui.adsabs.harvard.edu/abs/2013Sci...342..331H}{,
  331}}

\bibitem[{{Kamiaka} {et~al.}(2018){Kamiaka}, {Benomar}, \&
  {Suto}}]{kamiaka_reliability_2018}
{Kamiaka}, S., {Benomar}, O., \& {Suto}, Y. 2018,
  {\mhref{http://doi.org/10.1093/mnras/sty1358}{\mnras}},
  {\href{https://ui.adsabs.harvard.edu/abs/2018MNRAS.479..391K}{479}}{\href{https://ui.adsabs.harvard.edu/abs/2018MNRAS.479..391K}{,
  391}}

\bibitem[{{Koposov} {et~al.}(2023){Koposov}, {Speagle}, {Barbary}, {Ashton},
  {Bennett}, {Buchner}, {Scheffler}, {Cook}, {Talbot}, {Guillochon},
  {Cubillos}, {Asensio Ramos}, {Johnson}, {Lang}, {Ilya}, {Dartiailh}, {Nitz},
  {McCluskey}, \& {Archibald}}]{dynesty}
{Koposov}, S., {Speagle}, J., {Barbary}, K., {et~al.} 2023,
  {joshspeagle/dynesty: v2.1.3}, v2.1.3,  Zenodo,
  \dodoi{10.5281/zenodo.8408702}

\bibitem[{{Lai}(2012)}]{lai_tidal_2012}
{Lai}, D. 2012,
  {\mhref{http://doi.org/10.1111/j.1365-2966.2012.20893.x}{\mnras}},
  {\href{https://ui.adsabs.harvard.edu/abs/2012MNRAS.423..486L}{423}}{\href{https://ui.adsabs.harvard.edu/abs/2012MNRAS.423..486L}{,
  486}}

\bibitem[{{Landau} \& {Lifshitz}(1965)}]{landau_quantum_1965}
{Landau}, L.~D., \& {Lifshitz}, E.~M. 1965, {Quantum Mechanics. Nonrelativistic
  theory} (Oxford: Pergamon Press)

\bibitem[{{Li} {et~al.}(2024){Li}, {Deheuvels}, \&
  {Ballot}}]{li_asteroseismic_2024}
{Li}, G., {Deheuvels}, S., \& {Ballot}, J. 2024,
  {\mhref{http://doi.org/10.1051/0004-6361/202449882}{\aap}},
  {\href{https://ui.adsabs.harvard.edu/abs/2024A&A...688A.184L}{688}}{\href{https://ui.adsabs.harvard.edu/abs/2024A&A...688A.184L}{,
  A184}}

\bibitem[{{Li} {et~al.}(2014){Li}, {Naoz}, {Valsecchi}, {Johnson}, \&
  {Rasio}}]{li_dynamics_2014}
{Li}, G., {Naoz}, S., {Valsecchi}, F., {Johnson}, J.~A., \& {Rasio}, F.~A.
  2014, {\mhref{http://doi.org/10.1088/0004-637X/794/2/131}{\apj}},
  {\href{https://ui.adsabs.harvard.edu/abs/2014ApJ...794..131L}{794}}{\href{https://ui.adsabs.harvard.edu/abs/2014ApJ...794..131L}{,
  131}}

\bibitem[{{Lu} {et~al.}(2023){Lu}, {Rein}, {Tamayo}, {Hadden}, {Mardling},
  {Millholland}, \& {Laughlin}}]{lu_selfconsistent_2023}
{Lu}, T., {Rein}, H., {Tamayo}, D., {et~al.} 2023,
  {\mhref{http://doi.org/10.3847/1538-4357/acc06d}{\apj}},
  {\href{https://ui.adsabs.harvard.edu/abs/2023ApJ...948...41L}{948}}{\href{https://ui.adsabs.harvard.edu/abs/2023ApJ...948...41L}{,
  41}}

\bibitem[{{Lynden-Bell} \& {Ostriker}(1967)}]{lyndenbell_stability_1967}
{Lynden-Bell}, D., \& {Ostriker}, J.~P. 1967,
  {\mhref{http://doi.org/10.1093/mnras/136.3.293}{\mnras}},
  {\href{https://ui.adsabs.harvard.edu/abs/1967MNRAS.136..293L}{136}}{\href{https://ui.adsabs.harvard.edu/abs/1967MNRAS.136..293L}{,
  293}}

\bibitem[{Mathur {et~al.}(2019)Mathur, Santos, \& García}]{kepseismic3}
Mathur, S., Santos, {\^{A}ngela}., \& García, R.~A. 2019, Kepler Light Curves
  Optimized For Asteroseismology ("KEPSEISMIC"),  STScI/MAST,
  \dodoi{10.17909/T9-MRPW-GC07}

\bibitem[{{Maxted} {et~al.}(2015){Maxted}, {Serenelli}, \&
  {Southworth}}]{maxted_comparison_2015}
{Maxted}, P.~F.~L., {Serenelli}, A.~M., \& {Southworth}, J. 2015,
  {\mhref{http://doi.org/10.1051/0004-6361/201525774}{\aap}},
  {\href{https://ui.adsabs.harvard.edu/abs/2015A&A...577A..90M}{577}}{\href{https://ui.adsabs.harvard.edu/abs/2015A&A...577A..90M}{,
  A90}}

\bibitem[{{McLaughlin}(1924)}]{mclaughlin_results_1924}
{McLaughlin}, D.~B. 1924, {\mhref{http://doi.org/10.1086/142826}{\apj}},
  {\href{https://ui.adsabs.harvard.edu/abs/1924ApJ....60...22M}{60}}{\href{https://ui.adsabs.harvard.edu/abs/1924ApJ....60...22M}{,
  22}}

\bibitem[{{Millholland} \& {Spalding}(2020)}]{millholland_formation_2020}
{Millholland}, S.~C., \& {Spalding}, C. 2020,
  {\mhref{http://doi.org/10.3847/1538-4357/abc4e5}{\apj}},
  {\href{https://ui.adsabs.harvard.edu/abs/2020ApJ...905...71M}{905}}{\href{https://ui.adsabs.harvard.edu/abs/2020ApJ...905...71M}{,
  71}}

\bibitem[{{Mosser} {et~al.}(2018){Mosser}, {Gehan}, {Belkacem}, {Samadi},
  {Michel}, \& {Goupil}}]{mosser_period_2018}
{Mosser}, B., {Gehan}, C., {Belkacem}, K., {et~al.} 2018,
  {\mhref{http://doi.org/10.1051/0004-6361/201832777}{\aap}},
  {\href{https://ui.adsabs.harvard.edu/abs/2018A&A...618A.109M}{618}}{\href{https://ui.adsabs.harvard.edu/abs/2018A&A...618A.109M}{,
  A109}}

\bibitem[{{Nielsen} {et~al.}(2021){Nielsen}, {Davies}, {Ball}, {Lyttle}, {Li},
  {Hall}, {Chaplin}, {Gaulme}, {Carboneau}, {Ong}, {Garc{\'\i}a}, {Mosser},
  {Roxburgh}, {Corsaro}, {Benomar}, {Moya}, \& {Lund}}]{pbjam}
{Nielsen}, M.~B., {Davies}, G.~R., {Ball}, W.~H., {et~al.} 2021,
  {\mhref{http://doi.org/10.3847/1538-3881/abcd39}{\aj}},
  {\href{https://ui.adsabs.harvard.edu/abs/2021AJ....161...62N}{161}}{\href{https://ui.adsabs.harvard.edu/abs/2021AJ....161...62N}{,
  62}}

\bibitem[{{Ong}(2024)}]{ong_redgiant_2024}
{Ong}, J.~M.~J. 2024, {\mhref{http://doi.org/10.3847/1538-4357/ad0cac}{\apj}},
  {\href{https://ui.adsabs.harvard.edu/abs/2024ApJ...960....2O}{960}}{\href{https://ui.adsabs.harvard.edu/abs/2024ApJ...960....2O}{,
  2}}

\bibitem[{{Ong} \& {Basu}(2020)}]{ong_semianalytic_2020}
{Ong}, J.~M.~J., \& {Basu}, S. 2020,
  {\mhref{http://doi.org/10.3847/1538-4357/ab9ffb}{\apj}},
  {\href{https://ui.adsabs.harvard.edu/abs/2020ApJ...898..127O}{898}}{\href{https://ui.adsabs.harvard.edu/abs/2020ApJ...898..127O}{,
  127}}

\bibitem[{{Ong} {et~al.}(2022){Ong}, {Bugnet}, \& {Basu}}]{ong_rotation_2022}
{Ong}, J.~M.~J., {Bugnet}, L., \& {Basu}, S. 2022,
  {\mhref{http://doi.org/10.3847/1538-4357/ac97e7}{\apj}},
  {\href{https://ui.adsabs.harvard.edu/abs/2022ApJ...940...18O}{940}}{\href{https://ui.adsabs.harvard.edu/abs/2022ApJ...940...18O}{,
  18}}

\bibitem[{{Ong} \& {Gehan}(2023)}]{ong_rotation_2023}
{Ong}, J.~M.~J., \& {Gehan}, C. 2023,
  {\mhref{http://doi.org/10.3847/1538-4357/acbf2f}{\apj}},
  {\href{https://ui.adsabs.harvard.edu/abs/2023ApJ...946...92O}{946}}{\href{https://ui.adsabs.harvard.edu/abs/2023ApJ...946...92O}{,
  92}}

\bibitem[{{Ong} {et~al.}(2024{\natexlab{a}}){Ong}, {Nielsen}, {Hatt}, \&
  {Davies}}]{ong_reggae_2024}
{Ong}, J. M.~J., {Nielsen}, M., {Hatt}, E., \& {Davies}, G. 2024{\natexlab{a}},
  {\mhref{http://doi.org/10.21105/joss.06588}{JOSS}},
  {\href{https://ui.adsabs.harvard.edu/abs/2024JOSS....9.6588O}{9}}{\href{https://ui.adsabs.harvard.edu/abs/2024JOSS....9.6588O}{,
  6588}}

\bibitem[{{Ong} {et~al.}(2024{\natexlab{b}}){Ong}, {Hon}, {Soares-Furtado},
  {Stephan}, {van Saders}, {Tayar}, {Shappee}, {Hey}, {Cao}, {Y{\i}ld{\i}z},
  {Orhan}, {{\"O}rtel}, {Montet}, {Holoien}, {Bland-Hawthorn}, {Buder}, {De
  Silva}, {Freeman}, {Martell}, {Lewis}, {Sharma}, \& {Stello}}]{ong_zvrk_2024}
{Ong}, J.~M.~J., {Hon}, M. T.~Y., {Soares-Furtado}, M., {et~al.}
  2024{\natexlab{b}}, {\mhref{http://doi.org/10.3847/1538-4357/ad2ea2}{\apj}},
  {\href{https://ui.adsabs.harvard.edu/abs/2024ApJ...966...42O}{966}}{\href{https://ui.adsabs.harvard.edu/abs/2024ApJ...966...42O}{,
  42}}

\bibitem[{{Otor} {et~al.}(2016){Otor}, {Montet}, {Johnson}, {Charbonneau},
  {Collier-Cameron}, {Howard}, {Isaacson}, {Latham}, {Lopez-Morales}, {Lovis},
  {Mayor}, {Micela}, {Molinari}, {Pepe}, {Piotto}, {Phillips}, {Queloz},
  {Rice}, {Sasselov}, {S{\'e}gransan}, {Sozzetti}, {Udry}, \&
  {Watson}}]{otor_k56_2016}
{Otor}, O.~J., {Montet}, B.~T., {Johnson}, J.~A., {et~al.} 2016,
  {\mhref{http://doi.org/10.3847/0004-6256/152/6/165}{\aj}},
  {\href{https://ui.adsabs.harvard.edu/abs/2016AJ....152..165O}{152}}{\href{https://ui.adsabs.harvard.edu/abs/2016AJ....152..165O}{,
  165}}

\bibitem[{{Patel} \& {Penev}(2022)}]{patel_constraining_2022}
{Patel}, R., \& {Penev}, K. 2022,
  {\mhref{http://doi.org/10.1093/mnras/stac203}{\mnras}},
  {\href{https://ui.adsabs.harvard.edu/abs/2022MNRAS.512.3651P}{512}}{\href{https://ui.adsabs.harvard.edu/abs/2022MNRAS.512.3651P}{,
  3651}}

\bibitem[{{Paxton} {et~al.}(2011){Paxton}, {Bildsten}, {Dotter}, {Herwig},
  {Lesaffre}, \& {Timmes}}]{mesa_paper_1}
{Paxton}, B., {Bildsten}, L., {Dotter}, A., {et~al.} 2011,
  {\mhref{http://doi.org/10.1088/0067-0049/192/1/3}{\apjs}},
  {\href{https://ui.adsabs.harvard.edu/abs/2011ApJS..192....3P}{192}}{\href{https://ui.adsabs.harvard.edu/abs/2011ApJS..192....3P}{,
  3}}

\bibitem[{{Paxton} {et~al.}(2013){Paxton}, {Cantiello}, {Arras}, {Bildsten},
  {Brown}, {Dotter}, {Mankovich}, {Montgomery}, {Stello}, {Timmes}, \&
  {Townsend}}]{mesa_paper_2}
{Paxton}, B., {Cantiello}, M., {Arras}, P., {et~al.} 2013,
  {\mhref{http://doi.org/10.1088/0067-0049/208/1/4}{\apjs}},
  {\href{https://ui.adsabs.harvard.edu/abs/2013ApJS..208....4P}{208}}{\href{https://ui.adsabs.harvard.edu/abs/2013ApJS..208....4P}{,
  4}}

\bibitem[{{Paxton} {et~al.}(2015){Paxton}, {Marchant}, {Schwab}, {Bauer},
  {Bildsten}, {Cantiello}, {Dessart}, {Farmer}, {Hu}, {Langer}, {Townsend},
  {Townsley}, \& {Timmes}}]{mesa_paper_3}
{Paxton}, B., {Marchant}, P., {Schwab}, J., {et~al.} 2015,
  {\mhref{http://doi.org/10.1088/0067-0049/220/1/15}{\apjs}},
  {\href{https://ui.adsabs.harvard.edu/abs/2015ApJS..220...15P}{220}}{\href{https://ui.adsabs.harvard.edu/abs/2015ApJS..220...15P}{,
  15}}

\bibitem[{{Paxton} {et~al.}(2018){Paxton}, {Schwab}, {Bauer}, {Bildsten},
  {Blinnikov}, {Duffell}, {Farmer}, {Goldberg}, {Marchant}, {Sorokina},
  {Thoul}, {Townsend}, \& {Timmes}}]{mesa_paper_4}
{Paxton}, B., {Schwab}, J., {Bauer}, E.~B., {et~al.} 2018,
  {\mhref{http://doi.org/10.3847/1538-4365/aaa5a8}{\apjs}},
  {\href{https://ui.adsabs.harvard.edu/abs/2018ApJS..234...34P}{234}}{\href{https://ui.adsabs.harvard.edu/abs/2018ApJS..234...34P}{,
  34}}

\bibitem[{{Paxton} {et~al.}(2019){Paxton}, {Smolec}, {Schwab}, {Gautschy},
  {Bildsten}, {Cantiello}, {Dotter}, {Farmer}, {Goldberg}, {Jermyn}, {Kanbur},
  {Marchant}, {Thoul}, {Townsend}, {Wolf}, {Zhang}, \& {Timmes}}]{mesa_paper_5}
{Paxton}, B., {Smolec}, R., {Schwab}, J., {et~al.} 2019,
  {\mhref{http://doi.org/10.3847/1538-4365/ab2241}{\apjs}},
  {\href{https://ui.adsabs.harvard.edu/abs/2019ApJS..243...10P}{243}}{\href{https://ui.adsabs.harvard.edu/abs/2019ApJS..243...10P}{,
  10}}

\bibitem[{{Pires} {et~al.}(2015){Pires}, {Mathur}, {Garc{\'\i}a}, {Ballot},
  {Stello}, \& {Sato}}]{kepseismic2}
{Pires}, S., {Mathur}, S., {Garc{\'\i}a}, R.~A., {et~al.} 2015,
  {\mhref{http://doi.org/10.1051/0004-6361/201322361}{\aap}},
  {\href{https://ui.adsabs.harvard.edu/abs/2015A&A...574A..18P}{574}}{\href{https://ui.adsabs.harvard.edu/abs/2015A&A...574A..18P}{,
  A18}}

\bibitem[{{Rein} \& {Liu}(2012)}]{rebound}
{Rein}, H., \& {Liu}, S.~F. 2012,
  {\mhref{http://doi.org/10.1051/0004-6361/201118085}{\aap}},
  {\href{https://ui.adsabs.harvard.edu/abs/2012A&A...537A.128R}{537}}{\href{https://ui.adsabs.harvard.edu/abs/2012A&A...537A.128R}{,
  A128}}

\bibitem[{{Rice} {et~al.}(2022){Rice}, {Wang}, \&
  {Laughlin}}]{rice_origins_2022}
{Rice}, M., {Wang}, S., \& {Laughlin}, G. 2022,
  {\mhref{http://doi.org/10.3847/2041-8213/ac502d}{\apjl}},
  {\href{https://ui.adsabs.harvard.edu/abs/2022ApJ...926L..17R}{926}}{\href{https://ui.adsabs.harvard.edu/abs/2022ApJ...926L..17R}{,
  L17}}

\bibitem[{{Rogers} \& {Lin}(2013)}]{rogers_lin_2013}
{Rogers}, T.~M., \& {Lin}, D.~N.~C. 2013,
  {\mhref{http://doi.org/10.1088/2041-8205/769/1/L10}{\apjl}},
  {\href{https://ui.adsabs.harvard.edu/abs/2013ApJ...769L..10R}{769}}{\href{https://ui.adsabs.harvard.edu/abs/2013ApJ...769L..10R}{,
  L10}}

\bibitem[{{Rogers} {et~al.}(2012){Rogers}, {Lin}, \&
  {Lau}}]{rogers_internal_2012}
{Rogers}, T.~M., {Lin}, D.~N.~C., \& {Lau}, H.~H.~B. 2012,
  {\mhref{http://doi.org/10.1088/2041-8205/758/1/L6}{\apjl}},
  {\href{https://ui.adsabs.harvard.edu/abs/2012ApJ...758L...6R}{758}}{\href{https://ui.adsabs.harvard.edu/abs/2012ApJ...758L...6R}{,
  L6}}

\bibitem[{{Rogers} {et~al.}(2013){Rogers}, {Lin}, {McElwaine}, \&
  {Lau}}]{rogers_internal_2013}
{Rogers}, T.~M., {Lin}, D.~N.~C., {McElwaine}, J.~N., \& {Lau}, H.~H.~B. 2013,
  {\mhref{http://doi.org/10.1088/0004-637X/772/1/21}{\apj}},
  {\href{https://ui.adsabs.harvard.edu/abs/2013ApJ...772...21R}{772}}{\href{https://ui.adsabs.harvard.edu/abs/2013ApJ...772...21R}{,
  21}}

\bibitem[{{Rossiter}(1924)}]{rossiter_detection_1924}
{Rossiter}, R.~A. 1924, {\mhref{http://doi.org/10.1086/142825}{\apj}},
  {\href{https://ui.adsabs.harvard.edu/abs/1924ApJ....60...15R}{60}}{\href{https://ui.adsabs.harvard.edu/abs/1924ApJ....60...15R}{,
  15}}

\bibitem[{{Saunders} {et~al.}(2023){Saunders}, {Ong}, \&
  {Basu}}]{saunders_evolutionary_2023}
{Saunders}, D.~P., {Ong}, J.~M.~J., \& {Basu}, S. 2023,
  {\mhref{http://doi.org/10.3847/1538-4357/acbdf3}{\apj}},
  {\href{https://ui.adsabs.harvard.edu/abs/2023ApJ...947...22S}{947}}{\href{https://ui.adsabs.harvard.edu/abs/2023ApJ...947...22S}{,
  22}}

\bibitem[{{Saunders} {et~al.}(2024){Saunders}, {Grunblatt}, {Chontos}, {Dai},
  {Huber}, {Zhang}, {Stef{\'a}nsson}, {van Saders}, {Winn}, {Hey}, {Howard},
  {Fulton}, {Isaacson}, {Beard}, {Giacalone}, {Van Zandt}, {Murphey}, {Rice},
  {Blunt}, {Turtelboom}, {Dalba}, {Lubin}, {Brinkman}, {Louden}, {Page},
  {Watkins}, {Collins}, {Stockdale}, {Tan}, {Schwarz}, {Massey}, {Howell},
  {Vanderburg}, {Ricker}, {Jenkins}, {Seager}, {Christiansen}, {Daylan},
  {Falk}, {Brodheim}, {Gibson}, {Hill}, {Holden}, {Householder}, {Kaye},
  {Laher}, {Lanclos}, {Petigura}, {Roy}, {Rubenzahl}, {Schwab}, {Shaum},
  {Sirk}, {Smith}, {Walawender}, \& {Yeh}}]{saunders_obliquity_2024}
{Saunders}, N., {Grunblatt}, S.~K., {Chontos}, A., {et~al.} 2024,
  {\mhref{http://doi.org/10.3847/1538-3881/ad543b}{\aj}},
  {\href{https://ui.adsabs.harvard.edu/abs/2024AJ....168...81S}{168}}{\href{https://ui.adsabs.harvard.edu/abs/2024AJ....168...81S}{,
  81}}

\bibitem[{Schiesser(2012)}]{schiesser_numerical_2012}
Schiesser, W.~E. 2012, The numerical method of lines: integration of partial
  differential equations (San Diego: Academic Press)

\bibitem[{{Shibahashi}(1979)}]{shibahashi_modal_1979}
{Shibahashi}, H. 1979, \pasj,
  {\href{https://ui.adsabs.harvard.edu/abs/1979PASJ...31...87S}{31}}{\href{https://ui.adsabs.harvard.edu/abs/1979PASJ...31...87S}{,
  87}}

\bibitem[{{Shibahashi} \& {Saio}(1985)}]{shibahashi_rotational_1985}
{Shibahashi}, H., \& {Saio}, H. 1985, \pasj,
  {\href{https://ui.adsabs.harvard.edu/abs/1985PASJ...37..601S}{37}}{\href{https://ui.adsabs.harvard.edu/abs/1985PASJ...37..601S}{,
  601}}

\bibitem[{{Sikdar} \& {Dumberry}(2023)}]{sikdar_differential_2023}
{Sikdar}, B., \& {Dumberry}, M. 2023,
  {\mhref{http://doi.org/10.1016/j.pepi.2023.107022}{Physics of the Earth and
  Planetary Interiors}},
  {\href{https://ui.adsabs.harvard.edu/abs/2023PEPI..33907022S}{339}}{\href{https://ui.adsabs.harvard.edu/abs/2023PEPI..33907022S}{,
  107022}}

\bibitem[{{Spalding} \& {Batygin}(2014)}]{spalding_early_2014}
{Spalding}, C., \& {Batygin}, K. 2014,
  {\mhref{http://doi.org/10.1088/0004-637X/790/1/42}{\apj}},
  {\href{https://ui.adsabs.harvard.edu/abs/2014ApJ...790...42S}{790}}{\href{https://ui.adsabs.harvard.edu/abs/2014ApJ...790...42S}{,
  42}}

\bibitem[{{Spalding} \& {Batygin}(2016)}]{spalding_misalignment_2016}
---. 2016, {\mhref{http://doi.org/10.3847/0004-637X/830/1/5}{\apj}},
  {\href{https://ui.adsabs.harvard.edu/abs/2016ApJ...830....5S}{830}}{\href{https://ui.adsabs.harvard.edu/abs/2016ApJ...830....5S}{,
  5}}

\bibitem[{{Spalding} \& {Millholland}(2020)}]{spalding_oblateness_2020}
{Spalding}, C., \& {Millholland}, S.~C. 2020,
  {\mhref{http://doi.org/10.3847/1538-3881/aba629}{\aj}},
  {\href{https://ui.adsabs.harvard.edu/abs/2020AJ....160..105S}{160}}{\href{https://ui.adsabs.harvard.edu/abs/2020AJ....160..105S}{,
  105}}

\bibitem[{{Steffen} {et~al.}(2013){Steffen}, {Fabrycky}, {Agol}, {Ford},
  {Morehead}, {Cochran}, {Lissauer}, {Adams}, {Borucki}, {Bryson}, {Caldwell},
  {Dupree}, {Jenkins}, {Robertson}, {Rowe}, {Seader}, {Thompson}, \&
  {Twicken}}]{steffen_k56_2013}
{Steffen}, J.~H., {Fabrycky}, D.~C., {Agol}, E., {et~al.} 2013,
  {\mhref{http://doi.org/10.1093/mnras/sts090}{\mnras}},
  {\href{https://ui.adsabs.harvard.edu/abs/2013MNRAS.428.1077S}{428}}{\href{https://ui.adsabs.harvard.edu/abs/2013MNRAS.428.1077S}{,
  1077}}

\bibitem[{{Stephan} {et~al.}(2018){Stephan}, {Naoz}, \&
  {Gaudi}}]{stephan_destroyers_2018}
{Stephan}, A.~P., {Naoz}, S., \& {Gaudi}, B.~S. 2018,
  {\mhref{http://doi.org/10.3847/1538-3881/aad6e5}{\aj}},
  {\href{https://ui.adsabs.harvard.edu/abs/2018AJ....156..128S}{156}}{\href{https://ui.adsabs.harvard.edu/abs/2018AJ....156..128S}{,
  128}}

\bibitem[{{Stephan} {et~al.}(2021){Stephan}, {Naoz}, \&
  {Gaudi}}]{stephan_giant_2021}
---. 2021, {\mhref{http://doi.org/10.3847/1538-4357/ac22a9}{\apj}},
  {\href{https://ui.adsabs.harvard.edu/abs/2021ApJ...922....4S}{922}}{\href{https://ui.adsabs.harvard.edu/abs/2021ApJ...922....4S}{,
  4}}

\bibitem[{{Su} \& {Lai}(2022)}]{su_dynamics_2022}
{Su}, Y., \& {Lai}, D. 2022,
  {\mhref{http://doi.org/10.1093/mnras/stab3172}{\mnras}},
  {\href{https://ui.adsabs.harvard.edu/abs/2022MNRAS.509.3301S}{509}}{\href{https://ui.adsabs.harvard.edu/abs/2022MNRAS.509.3301S}{,
  3301}}

\bibitem[{{Tamayo} {et~al.}(2020){Tamayo}, {Rein}, {Shi}, \&
  {Hernandez}}]{reboundx}
{Tamayo}, D., {Rein}, H., {Shi}, P., \& {Hernandez}, D.~M. 2020,
  {\mhref{http://doi.org/10.1093/mnras/stz2870}{\mnras}},
  {\href{https://ui.adsabs.harvard.edu/abs/2020MNRAS.491.2885T}{491}}{\href{https://ui.adsabs.harvard.edu/abs/2020MNRAS.491.2885T}{,
  2885}}

\bibitem[{{Tayar} {et~al.}(2022){Tayar}, {Moyano}, {Soares-Furtado}, {Escorza},
  {Joyce}, {Martell}, {Garc{\'\i}a}, {Breton}, {Mathis}, {Mathur}, {Delsanti},
  {Kiefer}, {Reffert}, {Bowman}, {Van Reeth}, {Shetye}, {Gehan}, \&
  {Grunblatt}}]{tayar_spinning_2022}
{Tayar}, J., {Moyano}, F.~D., {Soares-Furtado}, M., {et~al.} 2022,
  {\mhref{http://doi.org/10.3847/1538-4357/ac9312}{\apj}},
  {\href{https://ui.adsabs.harvard.edu/abs/2022ApJ...940...23T}{940}}{\href{https://ui.adsabs.harvard.edu/abs/2022ApJ...940...23T}{,
  23}}

\bibitem[{{Tejada Arevalo} {et~al.}(2021){Tejada Arevalo}, {Winn}, \&
  {Anderson}}]{arevalo_further_2021}
{Tejada Arevalo}, R.~A., {Winn}, J.~N., \& {Anderson}, K.~R. 2021,
  {\mhref{http://doi.org/10.3847/1538-4357/ac1429}{\apj}},
  {\href{https://ui.adsabs.harvard.edu/abs/2021ApJ...919..138T}{919}}{\href{https://ui.adsabs.harvard.edu/abs/2021ApJ...919..138T}{,
  138}}

\bibitem[{{The pandas development Team}(2024)}]{pandas}
{The pandas development Team}. 2024, {pandas-dev/pandas: Pandas}, v2.2.3,
  Zenodo, \dodoi{10.5281/zenodo.3509134}

\bibitem[{{Townsend} \& {Teitler}(2013)}]{townsend_gyre_2013}
{Townsend}, R.~H.~D., \& {Teitler}, S.~A. 2013,
  {\mhref{http://doi.org/10.1093/mnras/stt1533}{\mnras}},
  {\href{https://ui.adsabs.harvard.edu/abs/2013MNRAS.435.3406T}{435}}{\href{https://ui.adsabs.harvard.edu/abs/2013MNRAS.435.3406T}{,
  3406}}

\bibitem[{{Triana} {et~al.}(2017){Triana}, {Corsaro}, {De Ridder}, {Bonanno},
  {P{\'e}rez Hern{\'a}ndez}, \& {Garc{\'\i}a}}]{triana_internal_2017}
{Triana}, S.~A., {Corsaro}, E., {De Ridder}, J., {et~al.} 2017,
  {\mhref{http://doi.org/10.1051/0004-6361/201629186}{\aap}},
  {\href{https://ui.adsabs.harvard.edu/abs/2017A&A...602A..62T}{602}}{\href{https://ui.adsabs.harvard.edu/abs/2017A&A...602A..62T}{,
  A62}}

\bibitem[{{Unno} {et~al.}(1989){Unno}, {Osaki}, {Ando}, {Saio}, \&
  {Shibahashi}}]{unno_nonradial_1989}
{Unno}, W., {Osaki}, Y., {Ando}, H., {Saio}, H., \& {Shibahashi}, H. 1989,
  {Nonradial oscillations of stars} (Tokyo: University of Tokyo Press)

\bibitem[{{van Saders} \& {Pinsonneault}(2013)}]{vansaders_fast_2013}
{van Saders}, J.~L., \& {Pinsonneault}, M.~H. 2013,
  {\mhref{http://doi.org/10.1088/0004-637X/776/2/67}{\apj}},
  {\href{https://ui.adsabs.harvard.edu/abs/2013ApJ...776...67V}{776}}{\href{https://ui.adsabs.harvard.edu/abs/2013ApJ...776...67V}{,
  67}}

\bibitem[{{Virtanen} {et~al.}(2020){Virtanen}, {Gommers}, {Oliphant},
  {Haberland}, {Reddy}, {Cournapeau}, {Burovski}, {Peterson}, {Weckesser},
  {Bright}, {van der Walt}, {Brett}, {Wilson}, {Millman}, {Mayorov}, {Nelson},
  {Jones}, {Kern}, {Larson}, {Carey}, {Polat}, {Feng}, {Moore}, {VanderPlas},
  {Laxalde}, {Perktold}, {Cimrman}, {Henriksen}, {Quintero}, {Harris},
  {Archibald}, {Ribeiro}, {Pedregosa}, {van Mulbregt}, \& {SciPy 1. 0
  Contributors}}]{scipy}
{Virtanen}, P., {Gommers}, R., {Oliphant}, T.~E., {et~al.} 2020,
  {\mhref{http://doi.org/10.1038/s41592-019-0686-2}{Nature Methods}},
  {\href{https://ui.adsabs.harvard.edu/abs/2020NatMe..17..261V}{17}}{\href{https://ui.adsabs.harvard.edu/abs/2020NatMe..17..261V}{,
  261}}

\bibitem[{{Vissapragada} {et~al.}(2022){Vissapragada}, {Chontos},
  {Greklek-McKeon}, {Knutson}, {Dai}, {P{\'e}rez Gonz{\'a}lez}, {Grunblatt},
  {Huber}, \& {Saunders}}]{vissapragada_tidal_2022}
{Vissapragada}, S., {Chontos}, A., {Greklek-McKeon}, M., {et~al.} 2022,
  {\mhref{http://doi.org/10.3847/2041-8213/aca47e}{\apjl}},
  {\href{https://ui.adsabs.harvard.edu/abs/2022ApJ...941L..31V}{941}}{\href{https://ui.adsabs.harvard.edu/abs/2022ApJ...941L..31V}{,
  L31}}

\bibitem[{{Weiss} {et~al.}(2024){Weiss}, {Isaacson}, {Howard}, {Fulton},
  {Petigura}, {Fabrycky}, {Jontof-Hutter}, {Steffen}, {Schlichting}, {Wright},
  {Beard}, {Brinkman}, {Chontos}, {Giacalone}, {Hill}, {Kosiarek},
  {MacDougall}, {Mo{\v{c}}nik}, {Polanski}, {Turtelboom}, {Tyler}, \& {Van
  Zandt}}]{weiss_kgps_2024}
{Weiss}, L.~M., {Isaacson}, H., {Howard}, A.~W., {et~al.} 2024,
  {\mhref{http://doi.org/10.3847/1538-4365/ad0cab}{\apjs}},
  {\href{https://ui.adsabs.harvard.edu/abs/2024ApJS..270....8W}{270}}{\href{https://ui.adsabs.harvard.edu/abs/2024ApJS..270....8W}{,
  8}}

\bibitem[{{Winn} {et~al.}(2010){Winn}, {Fabrycky}, {Albrecht}, \&
  {Johnson}}]{winn_hot_2010}
{Winn}, J.~N., {Fabrycky}, D., {Albrecht}, S., \& {Johnson}, J.~A. 2010,
  {\mhref{http://doi.org/10.1088/2041-8205/718/2/L145}{\apjl}},
  {\href{https://ui.adsabs.harvard.edu/abs/2010ApJ...718L.145W}{718}}{\href{https://ui.adsabs.harvard.edu/abs/2010ApJ...718L.145W}{,
  L145}}

\bibitem[{{Winn} \& {Fabrycky}(2015)}]{winn_occurence_2015}
{Winn}, J.~N., \& {Fabrycky}, D.~C. 2015,
  {\mhref{http://doi.org/10.1146/annurev-astro-082214-122246}{\araa}},
  {\href{https://ui.adsabs.harvard.edu/abs/2015ARA&A..53..409W}{53}}{\href{https://ui.adsabs.harvard.edu/abs/2015ARA&A..53..409W}{,
  409}}

\bibitem[{{Winn} \& {Holman}(2005)}]{winn_obliquity_2005}
{Winn}, J.~N., \& {Holman}, M.~J. 2005,
  {\mhref{http://doi.org/10.1086/432834}{\apjl}},
  {\href{https://ui.adsabs.harvard.edu/abs/2005ApJ...628L.159W}{628}}{\href{https://ui.adsabs.harvard.edu/abs/2005ApJ...628L.159W}{,
  L159}}

\bibitem[{{Yee} {et~al.}(2020){Yee}, {Winn}, {Knutson}, {Patra},
  {Vissapragada}, {Zhang}, {Holman}, {Shporer}, \& {Wright}}]{yee_orbit_2020}
{Yee}, S.~W., {Winn}, J.~N., {Knutson}, H.~A., {et~al.} 2020,
  {\mhref{http://doi.org/10.3847/2041-8213/ab5c16}{\apjl}},
  {\href{https://ui.adsabs.harvard.edu/abs/2020ApJ...888L...5Y}{888}}{\href{https://ui.adsabs.harvard.edu/abs/2020ApJ...888L...5Y}{,
  L5}}

\bibitem[{{Zanazzi} {et~al.}(2024){Zanazzi}, {Dewberry}, \&
  {Chiang}}]{zanazzi_damping_2024}
{Zanazzi}, J.~J., {Dewberry}, J., \& {Chiang}, E. 2024,
  {\mhref{http://doi.org/10.3847/2041-8213/ad4644}{\apjl}},
  {\href{https://ui.adsabs.harvard.edu/abs/2024ApJ...967L..29Z}{967}}{\href{https://ui.adsabs.harvard.edu/abs/2024ApJ...967L..29Z}{,
  L29}}

\bibitem[{{Zhang} {et~al.}(2021){Zhang}, {Weiss}, {Huber}, {Blunt}, {Chontos},
  {Fulton}, {Grunblatt}, {Howard}, {Isaacson}, {Kosiarek}, {Petigura},
  {Rosenthal}, \& {Rubenzahl}}]{zhang_long_2021}
{Zhang}, J., {Weiss}, L.~M., {Huber}, D., {et~al.} 2021,
  {\mhref{http://doi.org/10.3847/1538-3881/ac0634}{\aj}},
  {\href{https://ui.adsabs.harvard.edu/abs/2021AJ....162...89Z}{162}}{\href{https://ui.adsabs.harvard.edu/abs/2021AJ....162...89Z}{,
  89}}

\end{thebibliography}
